\documentclass[12pt]{article} 
\input{epsf.tex}
\usepackage{epsfig,ldd_art }
\newcommand{\One}{1\kern-4.5pt1}

\begin{document}
\vspace{-20pt}
\addtolength{\baselineskip}{0.20\baselineskip}

\begin{center}
{\Large{\bf Forces, fluxes and quasi-particles in hot QCD}}
\medskip

\centerline{\bf Chris P. Korthals Altes }

\centerline{\sl Centre Physique Th\'eorique au CNRS,}
\centerline{\sl Case 907, Campus de Luminy, F13288, Marseille, France}
\end{center}

\abstract{
These lectures start with a brief overview of salient features of the 
critical region of hot QCD. The main emphasis is on the accurate description
of static plasma observables by the well-known hierarchy of reduced actions combined with 3D simulations above $T\sim 2T_c$. A striking pattern emerges, put in perspective by completing the quasi-particle picture.}

\section{Introduction}\label{sec:intro}

This school gathers experimentalists and 
theorists in about equal numbers. And that is not more than natural: what else is physics than a
playground for experiment and theory? In fact, this is the  Galilean concept of physics and the  raison d'\^etre of this school!

So I will try to oblige and explain to you the little I came to know 
over the last  years about the plasma state of QCD. With the data from 
RHIC presently coming out, this is an ideal subject for such a mixed audience.   
Also, it ties in with the theoretical lectures given by Profs Strassler on flux, Shuryak on QCD at the transition point and Teper on lattice results.

One of the first duties of a theorist is being able to tell the experimentalist
what to compare his data  with. And the main part of this lecture will be 
dealing with exactly that: well above the critical temperature, say  two to three times its estimated value of 170 MeV, a theorist can predict with good accuracy
the behaviour of equilibrium properties of the plasma. 
The RHIC experimentalist may be disappointed: what he sees is a far cry from a plasma in equilibrium: only recently monojets, increasing proton to pion ratios as function of centrality
of the heavy ion collision have been measured and are according to some~\cite{ullie} a sign of a plasma formed at about twice the critical temperature. It may be only at ALICE that we will attain  temperatures like a GeV.  At any rate it is  beyond my competence to discuss the experimental signatures for the production of a plasma in equilibrium during  what may be coined
as a fleeting moment. In fact the creation of a plasma state at RHIC or ALICE
is sometimes compared to a ``little bang''. But the comparison is somewhat biased: whereas it is clear that the expansion of the universe was involving time scales much larger than the typical time scales for the QCD plasma to come to equilibrium, the time scale of the heavy ion colliders is down by roughly  the ratio of the Planck mass and the pion mass. So the reconstruction of the ``little bang'' from the data remains a daunting task.

So as a consolation of some sort: hot QCD in equilibrium may be useful for the cosmologist...
 
Despite recent  interesting developments in high density QCD I will limit myself to 
zero density.

In section (\ref{sec:qualitrans}) I will discuss a very simplified version of the QCD transition.
This to set  the stage for the formal developments based on QCD.
Then in section (\ref{sec:qcd}) QCD and its global symmetries and order parameters are discussed. Global symmetries are paramount in shaping the phase diagram of QCD. 
One of these global symmetries is explained in more detail in section (\ref{sec:canoniczn}).

The change  
 in the range of the  forces ~\cite{polyakov} from the hadron phase to the plasma phase is the subject of section (\ref{sec:forces}). 
The confining force in the hadron phase gets screened in the plasma phase.
In QED it is the Coulomb force that gets screened.

 In QED one can introduce Dirac monopoles as external sources. One expects no screening at all (after all Galilei observed sun spots,
witnesses of the long range of static magnetic fields in the sun's plasma).
In QCD monopole  sources are  screened for any temperature!  

In the next section the strategy for doing perturbative calculations  is set up. The strategy is the use of a hierarchy of effective actions. 

They are the electrostatic action, obtained from QCD by integrating out all hard modes of order T. 

The magnetostatic action
is obtained from the electrostatic action by integrating out all fields with  an electric screening  mass. This leaves us with a universal action (only depending through its coupling parameters on the original QCD action parameters). It is universal in its form, containing  the magnetic field strength and covariant derivatives thereof. Six years ago Shaposhnikov gave a set of lectures at this school
on this subject~\cite{shaposh}, and you are vividly advised to read those.
In the remaining part of the  lecture we will concentrate on developments in the last four years. The main topic is  the perturbation series
as produced by the reduced actions and how well it converges when confronted with non-perturbative lattice simulations. The results seem to indicate that the quasi-particle picture is still a good guide, provided we are willing to
accept not only the transverse non-static gluons but also the static transverse (magnetic) gluons  as quasi-particles.

 In section (\ref{sec:work}) follow a few examples: Wilson loop, pressure, Debye screening mass, magnetic screening mass and 't Hooft loop.  

In section (\ref{sec:quasihooft}) we come back to the gluonic quasi-particles
and show how they determine through their flux the behaviour of the spatial 
't Hooft loop. In particular, simple  scaling laws follow  from counting rules which are backed by perturbative calculation.  For spatial  Wilson loops, analytic calculations are not available. We hypothesize  soft  transverse gluons follow 
analogous counting rules, and arrive at a scaling law for the Wilson loop which is verified by lattice simulations.

\section{Qualitative picture of the transition}\label{sec:qualitrans}

\begin{figure}[htb]
\begin{center}
\epsfig{file=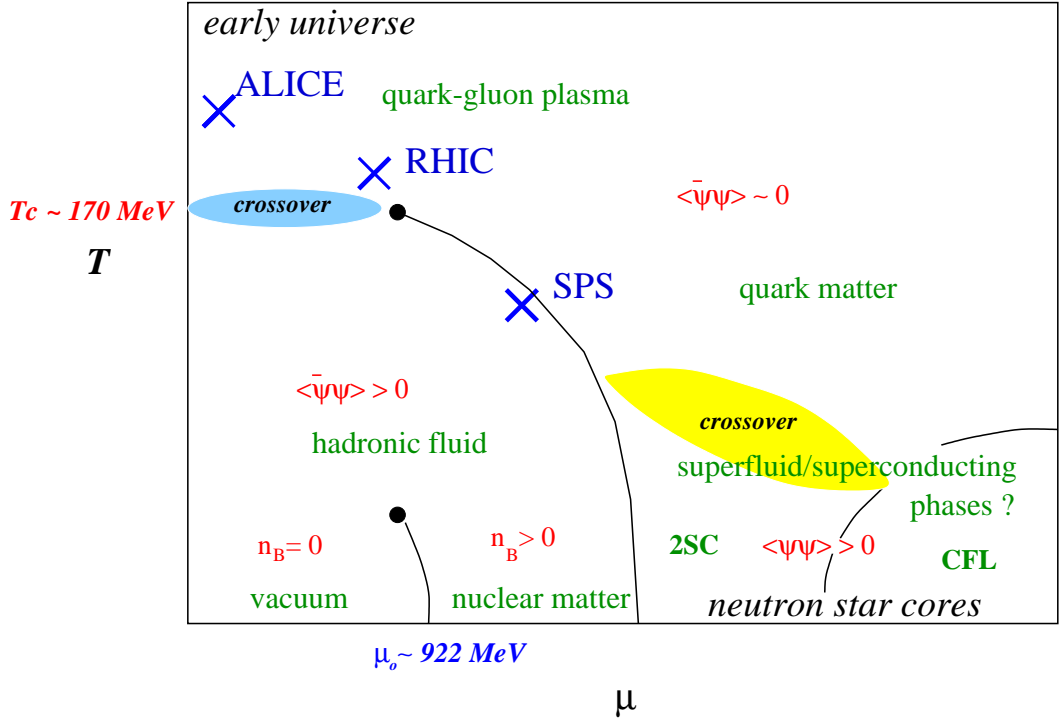, width=14cm}
\caption{Proposed phase diagram for QCD. 2SC and CFL refer
to the diquark condensates defined in ref.\cite{supercond}. From ref.\cite{hands}}
\label{fig:qcdphase}
\end{center}
\end{figure}

To know  better what we are talking about,  have a look at fig.(\ref{fig:qcdphase}).

%%%%%%%%%%%%%%%%%%%%%%%%%%%%%%%%%%%%%%%%%%%%%%%%%%%%%%%%%%%%%%%%%%%%%%%%%

It is a schematic phase diagram of QCD as function of temperature and nucleon density, or more precisely the nucleon chemical potential. At the origin we have a groundstate where the quarks and the antiquarks combine into Cooper pairs. This condensate of Cooper pairs is sensitive to temperature and chemical potential changes.  You see a familiar transition at zero temperature and chemical potential about 900 MeV: the formation of nuclear matter. For still higher chemical potential we get a a phenomenon
called Pauli blocking.  At high nucleon density the Pauli principle frustrates the formation of quark-antiquark pairs because the high density of nuclear matter renders all low lying quark levels occupied.  So the chiral quark condensate will diminish with growing density. On the other hand at very high densities the gauge coupling becomes so small that perturbation theory is valid. It tells us the pairing of quark-quark pairs is 
preferred. Then the Cooper instability changes the  groundstate  into a state of matter where we have instead of a condensate of paired electrons with electric superconductivity a condensate of
paired quarks  with colour superconductivity~\cite{supercond}. This phenomenon may take place in neutron star cores.   

Let us now increase the temperature T. On  the vertical axis at zero $\mu$
 probably a cross-over behaviour results for realistic quark masses. Cross-over behaviour means a gradual change of thermodynamic quantities, like pressure and internal energy. Numerical simulations for realistic quark masses are not yet decisive on this point. Some
lattice simulations of QCD~\cite{karschlect}~\cite{lainechemical} indicate a critical point for non-zero nucleon density.  At any rate, it is not excluded that the part of the phase diagram dubbed ``hadronic fluid'' is smoothly connected to the region where ALICE and RHIC will probe the phase diagram.

The continous lines show a first order transition. First order means that 
quantities like the energy density jump at the transition. At the endpoint, thermodynamics tells us the transition must be second or higher order. 
 In such points first derivatives of the free energy are still continuous (or even higher derivatives). A transition is often caused by a change in the way a global symmetry is realized\footnote{The Curie point in ferromagnetism is a transition where rotational symmetry is restored. Below the Curie point the ground
 state of the system is not rotationally invariant: there is a permanent magnetization.}. We will see more about that in the next sections.

So the diagram shows a rich variety in physics: collider experiments take place near the T-axis, cosmology on the T-axis, and astrophysics at low-T high density. That collider physics and cosmology have small density in common is a fortunate coincidence: one may have a direct bearing on the other and there is a rich litterature on this subject.
 ~From now on we concentrate on zero chemical potential.

\subsection{A simple model of the transition}

The simplest way to see there must be a transition is to take the bag model
of hadrons. Increase  the temperature up to energies $E$ on the order of the pion mass$\sim 140 MeV$. The Boltzmann probabilty $\exp{-{E\over T}}$ for thermal excitation  tells us a gas of relativistic pions has formed, with a Stefan-Boltzmann pressure:
\begin{equation}
p_L=3\mbox{x}{\pi^2\over {90}}T^4
\label{eq:pL}
\end{equation}

There is an isospin degeneracy factor 3 in this pressure at low $T$.

Similarly, coming in from temperatures $T$ much larger than  the pion mass, we can expect on the basis of asymptotic 
freedom a gas of free quarks and gluons. Taking the degrees of freedom into account (for a given number $N_f$ of flavours)  one finds:

\begin{equation}
p_H=p_{q\bar q}+p_{glue}
   =2\mbox{x}2\mbox{x}3\mbox{x}{7\over 8}N_f{\pi^2\over{90}}T^4+2\mbox{x}8\mbox{x}{\pi^2\over{90}}T^4
\label{eq:pH}
\end{equation}

Near the critical temperature the bag pressure $B$ of the hadrons 
is released and adds up to the pressure of the pionic gas. In other words, the individual hadron bags become one large bag. This is typically what percolation
is about. Percolation of the pions in the gas means they are starting to overlap with the result that quarks and gluons do not know anymore to what hadron they belong.  

So, when comparing the two pressures at $T_c$ one finds:
\begin{equation}
p_L(T_c)+B=p_H(T_c)
\label{eq:tcritbag}
\end{equation}

With a bag pressure $B\sim (200 \mbox{MeV})^4$ and $N_f=2$ one arrives at $T_c\sim 140 MeV$.

Here we suppose  the bag constant  not to  vary with temperature.
This means that the internal energy density is related to the pressure by  $\epsilon_L(T_c)=3p_L(T_c)$
whereas $\epsilon_H(T_c)=3p_H(T_c)$. So eq.(\ref{eq:tcritbag}) tells us
that the latent heat $\Delta\epsilon_c=\epsilon_H(T_c)-\epsilon_L(T_c)=3B$, a very strong first order 
transition indeed! Comparing the jump to the mean value $\epsilon_c$ one 
finds $\Delta\epsilon_c/ \epsilon_c=O(1)$.

Such spectacular jumps would leave their marks in distributions and correlations of the 
hadronic decay products. 

But we mentioned already a caveat: we supposed the bag constant not to vary
with T, and this was what made the transition first order.  

So the real question is: what does QCD say about the transition?    

\section{Global symmetries, order parameters and the phase transition in QCD}\label{sec:qcd}

The QCD action has as input parameters the experimental values of $\Lambda_{\overline{\rm MS}}$,the number of colours  $N$, the number of flavours $N_f$ and the masses $m_i$ of the quarks. Together with the 
Lagrangian:
\begin{equation}
{\cal L}_{QCD}={1\over 2}Tr F_{\mu\nu}F_{\mu\nu}+\sum_{i=1}^{N_f}\bar \psi_i(\gamma_{\mu}D_{\mu}+m_i)\psi_i
\label{eq:qcdaction}
\end{equation}

\noindent these input parameters describe all of hadron physics\footnote{ $D_{\mu}$ is the covariant derivative $\partial_{\mu}+igA_{\mu}$ and $F_{\mu\nu}=\partial_{\mu}A_{\nu}-\partial_{nu}A_{\mu}+ig[A_{\mu},A_{\nu}]$.}.

This Lagrangian is a strongly coupled system. Its particle spectrum consists of glueballs, and quark bound states. To test this Lagrangian, numerical simulations with a lattice version of QCD are done. This lattice version of the gauge field action is in terms of SU(N)  matrices $U(l)$ living on the links $l$ of the lattice. The links have length a, the cut-off in our theory. The field strength matrix $F_{\mu\nu}$ is replaced by the product of the link matrices on every plaquette $U(P)=\Pi_{l\in P}U(l)$. This product is the exponentiated flux through the plaquette:
\begin{equation}
U(P)=\Pi_{l\in P}U(l)=\exp{ia^2F_{\mu\nu}+...}
\label{eq:fluxcutoff}
\end{equation}
\noindent where we suppose the two sides of $P$ are in the $\mu\nu$ direction and where the dots mean higher derivatives.
 And the action density is replaced by:
\begin{equation}
Tr F_{\mu\nu}^2\rightarrow 1-{1\over N}Re TrU(P).
\label{eq:latticeaction}
\end{equation}

The lattice coupling  $\beta$ is related to the bare coupling $g$ by $\beta={2N\over {g^2}}$. For more details see the lecture notes by Prof. Teper in this volume.

 Global symmetries in QCD
depend on how the fermion masses are implemented. Two extremes determine qualitatively what we know for zero nucleon density. All quark masses zero or all
infinitely heavy. In the first case left handed quarks  $\psi_L$  and right handed quarks $\psi_R$ transform under the  symmetry group $SU(N_f)_L\mbox{x}SU(N_f)_R$. The Lagrangian stays invariant, but the symmetry is realized in the spontaneously broken mode: the left handed quark and its righthanded partner do couple in the real world through a term $\bar\psi_L\psi_R$, trace over flavour indices understood. And such a term is only invariant under a left handed symmetry operation in combination with the contragredient right handed
partner. This is the diagonal subgroup. The massless Goldstone bosons transform
as an adjoint multiplet under this group. Nature provides a non-zero vacuum expectation value of the 
left right coupling $$<\bar\psi_L \psi_R +\mbox{h.c.}>\sim (250 \mbox{MeV})^3.$$ It transforms non-trivially under
the group, so is a measure of the breaking of the symmetry. It is an order parameter. The Goldstone theorem assures then the existence of an adjoint multiplet of massless pseudoscalars.

We have left out the two $U(1)$ factors. One factor leaves the order parameter invariant and is connected to baryon number conservation. The other factor $U(1)_A$ transforms the condensate. But due to quantum corrections - the axial anomaly -  it is not a symmetry of the system and the corresponding Goldstone boson gets a mass due to the instanton mechanism~\cite{instanton}. 

\subsection{Universality}\label{subsec:universality}
The order parameter $\Phi$ of many statistical systems is zero above the critical
temperature $T_c$. Below $T_c$ it is non-zero, so its behaviour is non-analytic. Should it jump at $T_c$, the transition is called first order. If it is continuous but its first derivative jumps, it's called second order and so on. 
The order of the transition is the same for a whole class of statistical systems and this is called the universality class of the transition:
\begin{itemize}
\item
The order of the transition is determined by the symmetry and the dimensionality of the system as described by the order parameter.
\end{itemize}
So to know the order of the transition of QCD we just take 
the most general 3D
action consistent with the symmetries of QCD one can write down for the order parameter $\Phi_{ij}=\bar\psi_{L,i}\psi_{R,j}$. It is the following:

\begin{equation}
{\cal{L}}_{eff\chi}=Tr(\vec\partial\Phi^{\dagger}\vec\partial\Phi)+m^2Tr\Phi^{\dagger}\Phi+g(Tr\Phi^{\dagger}\Phi)^2+hTr(\Phi^{\dagger}\Phi)^2+ c\mbox{det}\Phi+\mbox{h.c}
\label{eq:wilczek}
\end{equation}

The critical behaviour of this action is the same as that of QCD according to universality. For $N_f=2$ the global symmetry is that of $O(4)\sim SU(2)\mbox{x}SU(2)$ and is known to be 2nd order. For $N_f=3$ the determinantal term drives it first order~\cite{wilczek}.
\newpage
\subsection{Z(N) symmetry} 
 Till now the masses of the quarks were zero. Let us  go to the other extreme: 
infinitely massive quarks. They leave us with only gluons as dynamical agents.

Then there is a global symmetry:  $Z(N)$ symmetry~\cite{thooft} with $N=3$ in case of QCD. $Z(N)$ stands for the subgroup of $SU(N)$ that commutes with all elements
of $SU(N)$. It consists of  $N$ matrices $z_k1_N$, $1_N$ the $NxN$ unit matrix, and $z_k=\exp{ik{2\pi\over N}}$, k=0,1,...N-1. For notational convenience we'll drop the unit matrix henceforth.

Where does this symmetry come from?  In contrast to chiral symmetry, 
Z(N) symmetry is not a symmetry acting on quantum states.
It is a symmetry of the free energy of the system expressed as a path integral. 

  To get this point we have to understand 
a basic fact about the description of static phenomena at finite temperature T.
Any observable $O$ has a thermal expectation value given by the Gibbs sum:
\begin{equation}
<O>_T=\mbox{Tr}~O\exp{(-H/T)}/\mbox{Tr}\exp{-H/T}
\label{eq:gibbs}
\end{equation}

The trace is over physical states only. Physical states are by definition gauge   invariant states, that is, invariant under gauge transformations that are regular in configuration space.  Of course only gauge invariant observables are admitted.
The factor $1/T$ in the Boltzmann factor is like an imaginary time span
 in a  quantum mechanical amplitude. The transcription to a path integral
is then straightforward~\cite{textbook}. The trace means the path integral will be periodic in this imaginary (``Euclidean'') time for bosons~\footnote{ And antiperiodic for fermions. In this formalism boundary conditions tell an important distinction between bosons and fermions: they tell
us the distinction in statistics!}.

An immediate consequence is the  transcription of Feynman rules. For finite temperature the  Feynman rules in Euclidean space undergo one single and simple change: instead of integration over energies, energies are now discrete because of the (anti)- periodicity.
For bosons they equal $\omega_n=2\pi Tn$, for fermions $\omega_n=2\pi(n+{1\over 2})T$. In both cases $n$ is integer. This change goes into the propagators, vertices and energy momentum conservation at the vertices.

Long  ago 't Hooft~\cite{thooft} realized that the periodicity in time does not necessarily mean you need
      gauge transformations to be periodic in time.
A gauge transformation can be periodic modulo a center group element $\exp{ik{2\pi\over N}}$ of the gauge group $SU(N)$. The gluon field being an adjoint does not feel any centergroup element. So action and measure of the path integral are insensitive to such a gauge transformation~\footnote{But quark fields are sensitive to the center group: antiperiodic boundary conditions are changed, and hence the statistics.}. 
   
So locally we have a gauge transformation. But observables that are non-local  over the whole periodicity range will 
feel a change, despite the fact that as observables they need to be gauge invariant against everywhere regular gauge transformations. 

A prime example, where  our discontinuous gauge transformation makes itself felt, is the 
Wilson line running in the periodic time direction:
\begin{equation}
P(A_0)(\vec x)={1\over N}\mbox{Tr}{\cal{P}}\exp{ig\int_0^{1\over T}d\tau A_0(\vec x,\tau)}
\label{eq:wilsonline}
\end {equation}

\noindent where the path ordering is defined by dividing the interval $[0,1/T]$
into a large number  $N_{\tau}$ of bits of length $\Delta\tau={1\over {N_{\tau}T}}$:
\begin{equation}
{\cal{P}}(A_0)=\lim_{N_{\tau}\rightarrow\infty} U(\tau=0,\Delta\tau)U(\tau=\Delta\tau ,2\Delta\tau)\ldots{.....}U(\tau={1\over T}-\Delta\tau,{1\over T})
\label{eq:wilsonlineordering}
\end {equation}
in an obvious notation. We have dropped the $\vec x$ dependence to avoid clutter in the formulae. Every factor is a string bit $U(\tau,\tau+\Delta\tau)=\exp{ig\Delta\tau A_0(\tau)}$. Every string bit in this product transforms under a gauge transform as $\Omega(\tau)U(\tau, \tau+\Delta\tau)\Omega^{\dagger}(\tau+\Delta\tau)+O({1\over{N_{\tau}^2}})$ . 
So periodic gauge transforms  transform the Wilson line like $\Omega(\tau=0){\cal{P}}(A_0)\Omega^{\dagger}(\tau=0)$. And so the trace is invariant. But a discontinuity
will multiply the Wilson line with the center group phase $z_k$, if
\begin{equation}
 \Omega_k(\tau={1\over T})=z_k^*\Omega_k(\tau=0) 
\label{eq:k}
\end{equation}
Note that a discontinuity other than the center group would be fixed at the time $\tau={1\over T}$ inside the trace of the Wilson line. Only the center group is a global group, i.e. it does not matter where in time the discontinuity was defined to be.

Although the Z(N) transformation  leaves the path integral, hence the free energy invariant, the question whether it commutes with the Hamiltonian of the system makes no sense. There is no such thing as a conserved charge.

This in contrast to canonical $Z(N)$  symmetries that do commute with the Hamiltonian. They can be broken at low temperature but not at high temperature as intuition has it. In  section (\ref{sec:canoniczn}) we illustrate this point in QCD.

\subsection{Wilson lines, Z(N) symmetry and the deconfining phase transition}\label{subsec:wilsonzndeconf}

The thermal average of the Wilson line is related to the free energy excess $\Delta F_{\psi}$ of a state with a very heavy test quark  $\psi_i$, averaged  over all gauge transforms of the state and averaged over  the $N$ colour indices $i$:  

\begin{equation}
\exp{-\Delta F_{\psi}/T}=\int DA{1\over N}\mbox{Tr}{\cal{P}}(A_0)\exp{-S(A)}/\int DA \exp{-S(A)}\equiv\langle P(A_0)\rangle.
\label{eq:excess}
\end{equation}

In appendix A we prove this relation. It is valid for any heavy point source in the fundamental representation. A source in any representation $r$ of the group will just change the representation of the Wilson line into $r$.

The thermal average of the Wilson line has been simulated and it is zero
at low temperatures, but at $T_c$ it rises abruptly to acquire the value 
$1$ at very high T.  A little thought makes this clear, because of its connection to the heavy fundamental charge.

The energy excess equals the 
energy of the fluxtube pointing from the test charge. As the flux cannot return to the test charge, the length of the fluxtube is typically the spatial size of the box. The energy equals the string tension times the length so is proportional to the size of the box.

    In the phase where the fluxlines are screened, this energy is finite and will become zero if screening is total.

However we have swept a problem under the rug, that of the short distance
effects on the self energy. They are still contained in the thermal average, eq.(\ref{eq:excess}),
and contribute in terms of the lattice cut-off $a$:
\begin{equation}
\Delta F_{\psi}\sim {1\over{Ta}}.
\label{eq:latticeself}
\end{equation}

For fixed temperature the lattice cut-off goes to zero exponentially fast
in the lattice coupling. On the other hand the inverse temperature equals 
size $N_{\tau}a$ with $N_{\tau}$ the number of lattice points in the Euclidean
time direction. So the free energy excess due to thermal effects is to be corrected for this short distance effect, and to do this is in practice quite intricate~\cite{karschlect}.

Let us illustrate how one determines the transition, say in SU(3). 
You can ask the question: what is the distribution of expectation values of the Wilson line $\overline{ P}$, averaged over the space volume of our box. Mathematically one asks the probability $E(\widetilde P)$ of a given value $\widetilde P$ of the line to occur:
\begin{equation}
E(\widetilde P)\sim\int DA \delta(\widetilde P-\overline{ P(A_0)}~)\exp{-S(A)}
\label{eq:effaction}
\end{equation}
Under a gauge transform $\Omega_k$ as in eq.(\ref{eq:k}) the measure and the action are invariant. Only the line average $\overline{ P(A_0)}$ picks up the factor $z_k$,
so 
\begin{equation}
E(\widetilde P)=E(z_k^*\widetilde P)
\end{equation}

\begin{figure}[tb]
\centerline{
\epsfxsize=14cm\epsfbox{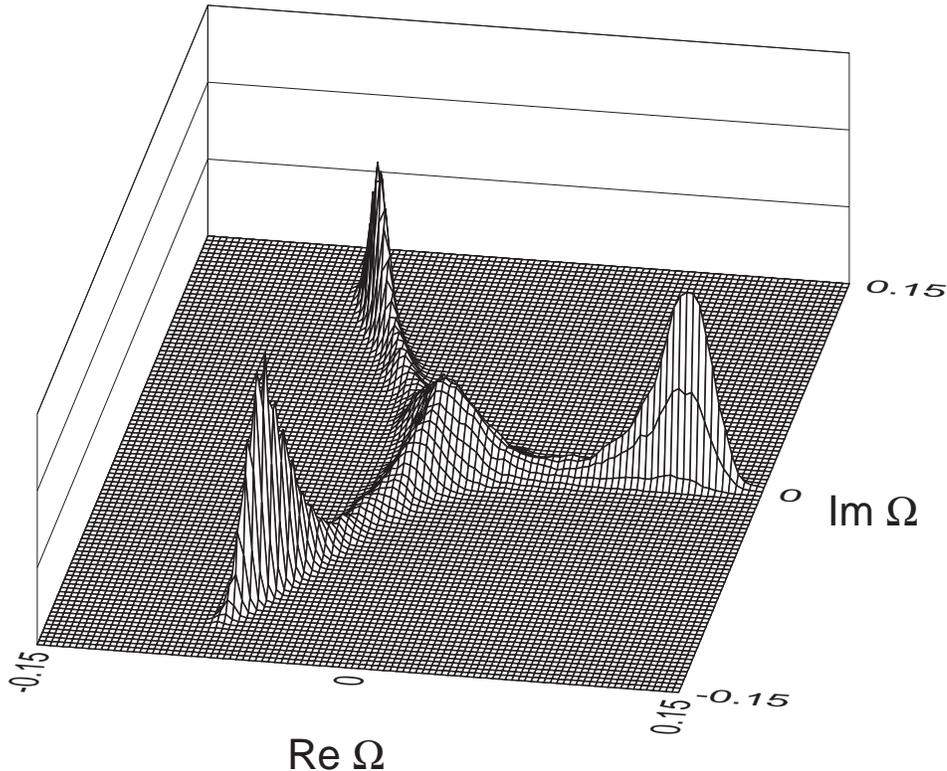}
}
\vspace{-0.7cm}
\caption{
Thermal Wilson line histogram in the SU(3) gauge theory at the deconfining
transition point obtained on
a $24^2\times36\times4$ lattice, QCDPAX collaboration\protect\cite{kanaya}.
}
\label{fig:PolyHstg}  

\end{figure}

For $SU(3)$ the distribution $E$ is shown in fig.(\ref{fig:PolyHstg}) at the transition temperature $T_d$. The three peaks at the center group values are equally populated, and the figure
clearly shows invariance under multiplication with $\exp{i{2\pi\over 3}}$. The central peak at $\widetilde P=0$ is a sign that the system likes to be in the hadron phase at the same time. This suggests coexistence of the hadron and deconfined phase at $T=T_d$.  At higher $T$ the central  peak disappears rapidly and we are left with the three peaks at the center group values. This is confirmed by perturbative calculation of the distribution~\cite{kuti}.

%%%%%%%%%%%%%%%%%%%%%%%%%%%%figure%%%%%%%%%%%%%%%%%%%%%%%%%%%%%%%%%%%%%%%%%%%%%
%%%%%%%%%%%%%%%%%%%%%%%%%%%%%%%%%%%%%%%%%%%%%%%%%%%%%%%%%%%%%%%%%%%%%%%%
So the behaviour of the Wilson line indicates that at low temperature the Z(3) symmetry is restored and broken at high temperature, at first sight counter-intuitive. It seems that at high T the Wilson line spins want to align.
This would be understandable if the surface tension between regions where the Wilson lines point in different Z(3) direction becomes very large at high T.
The surface tension has dimension $(\mbox{mass})^2$. In quarkless QCD there is no scale 
so the tension must be proportional to $T^2$. Hence alignment is energetically favourable at high T and the $Z(3)$ symmetry is spontaneously broken\footnote{
Note that the QCD scale $\Lambda$ was left out of the argument. Why were we allowed to do so? At high T the $\Lambda$ parameter is absorbed in the running coupling  and is nowhere else present in high T observables (see section(\ref{sec:reduction})).}.

\subsection{Z(N) universality}\label{subsec:znuniversality}

Let us now discuss universality in the context of the Wilson line.
The Wilson line is an order parameter and serves therefore to define the universality class~\footnote{For a thorough discussion see the lecture notes of Pisarski~\cite{pispis}.}. In analogy to the discussion of chiral symmetry, especially eq.(\ref{eq:wilczek}), we now look for a 3D action which has $Z(N)$ symmetry. The Wilson line $P(A_0)$ is now written as a complex number $p$. In any lattice point 
$\vec x$  we have an independent ``spin'' $p(\vec x)$, and it transforms under global $Z(N)$ as\begin{equation} 
p\rightarrow z^kp~\mbox{with}~ k=0,1,2,..,N-1~\mbox{and}~ z=\exp{i{2\pi\over N}}.
\label{eq:zntransfspin}
\end{equation}
So $p$ takes only values in the center group. 

Now it is an easy task to write down actions that are Z(N) invariant.
For N=2 it is the famous Ising model in 3D, that models spontaneous magnetization:
\begin{equation}
S_{N=2}=\beta\sum_{n.n.}\big(1-p(\vec x)p(\vec y)\big)
\label{eq:ising}
\end{equation}
The sum is  over all links that connect two neighbouring lattice points. 
One sums the Boltzmann factor $\exp{-S_N(\{p(\vec x)\})}$  over all configurations $\{p(\vec x)\}$  to get the free energy per lattice point ($V$ is the number of lattice points):
\begin{equation}
\exp{-\beta Vf_{N=2}(\beta)}=\sum_{\{p(\vec x)\}}\exp{-S_{N=2}(\{p(\vec x)\})}.
\end{equation} 

This gives the free energy and a transition
is found at a $\beta_c\sim O(1)$. Below this point the order parameter $\langle p\rangle=0$. This is understandable: at $\beta=0$ the relative sign of neighbouring spins does not matter in the Boltzmann factor, so disorder will prevail and no magnetization  $\langle p\rangle$ results.

 Above this point it starts to grow to attain the value 1 or -1 at large $\beta$. At large $\beta$ the spins at the end of any link align because that lowers the action. So the Boltzmann probability will be higher. Whether the resulting magnetization is positive or negative depends on the way we prepare the system. Even an infinitesimal applied magnetic field h ( in the guise of a term $h\sum_{\vec x}p(\vec x)$ in the action)
will decide about it. 

 One can induce a region of up-magnetization next to
a down-magnetization region by changing the interaction on the links that pierce the wall between the two regions. The change is from $1-p(\vec x)p(\vec y)$
to $1+p(\vec x)p(\vec y)$.  The effect of that change or ``twist'' is  for large $\beta$ that
the spins on such links will anti-align, because that optimizes the Boltzmann probability on such links. So this creates a domain wall around the twist  with a surface tension $\alpha(\beta)\ge 0~\mbox{as}~\beta\ge \beta_c$. An equivalent way to create the same domain wall is to fix spins to be up at one end of the volume, and down at the other end.  Exactly analogously, in gauge theory one can fix the thermal Wilson line  and compute a ``domain wall'' tension.  Alternatively, one can define a twist
in gauge theory most naturally in the lattice formulation.

The critical properties of the 3d Ising model have been well established, by numerical means, series expansions etc.. The transition is known to be second order.
So magnetization and surface tension go smoothly to zero above the critical
point. In particular $\alpha(\beta)\sim |\beta-\beta_c|^{2\nu}$, with $2\nu=1.26...$. 
And indeed  the corresponding transition in $SU(2)$ is second order, it turns out by lattice simulation. In a later section we shall see more manifestations of this universality for $SU(2)$, namely for the surface tension. The surface tension of a wall separating two regions where the thermal Wilson line has opposite signature has been simulated~\cite{defor} and is shown in fig. (\ref{fig:sigma}). Universality is well satisfied by the exponent. 

For SU(3) the spin action is Z(3) invariant and reads:
\begin{equation}
S_{N=3}=\beta\sum_{n.n.}\big(1-p(\vec x)p^*(\vec y) +\mbox{c.c.}\big) 
\end{equation}
The first term is obviously Z(3) invariant. The reality of the action reflects 
the charge conjugation invariance of the SU(3) theory. Charge conjugation guarantees that the average of $P$ and of $P^*$ is the same. And so does the reality
of the spin action for the average of $p$ and $p^*$.

Now the transition is first order for the spin system. 
And the SU(3) transition
is indeed first order, though weakly so~\cite{firstorder}. With weakly is meant that the ratio in the jump in energy over the energy is small, in contrast to what we found in the bag model before.

So universality seems to be well established for N=2 and 3. 

Not only the order of the transition but   quantities like exponents are identical according to universality. We will come back to those when 
discussing correlations.

\subsection{Universality for large number of colours}
Though QCD has three colours, GUT theories feature more, and it is therefore 
not academic to look at $N\ge 4$.
For N=4 and larger we find a  lack of predictivity. In fact  Z(N) spin models usually comprise different universality classes. Z(2) and Z(3) are rather the exception!  

The most general model with $Z(4)$ invariance has two parameters instead of one:
\begin{equation}
S_{N=4}=\beta\sum_{n.n.}\big(d_1\big(1- p(\vec x)p^*(\vec y)\big) +\mbox{c.c.}+d_2\big(1-(p(\vec x)p^*(\vec y))^2\big) \big).       
\label{eq:nis4}
\end{equation}
The normalization of $\beta$ is clearly a convention so we have indeed two physical degrees of freedom.

The two-dimensional phase diagram of this model is well known. It has first and second order transitions as one varies the ratio $d_1/d_2$. Setting $d_1=0$ the resulting model is  Ising-like because the remaining interaction term  fluctuates between $\pm 1$ and we saw before  it was second order.
Putting $d_1=d_2=d$ gives us an action with value $3\beta d$ if the spins are aligned, and $-\beta d$ if otherwise. This is a class of models - Potts models - which has  a first order transition from $N=3$ on. Depending on the ratio $d_1/d_2$ the order of the transition changes. In other words there are at least two   universality classes in Z(4) spin models , and the question is to which one the SU(N) gauge
theory belongs.

 Simulations of  SU(4)~\cite{ohta} gauge theory show a first order transition, and the same is true for $SU(6)$~\cite{teperhot}.

In conclusion: where it is defined, universality works well. For $N\ge 4$
we have to invent extra criteria to pinpoint the universality class in the
spin model.

%%%%%%%%%%%%%%%%%%%%%%%%%%%%%%%%%figure%%%%%%%%%%%%%%%%%%%%%%%%%%%%%%%%%%%%%%%%%
\begin{figure}[htb]
%\vspace{-0.6truecm}
%\vspace{9pt}
\begin{center}
\epsfig{bbllx=0,bblly=175,bburx=564,bbury=514,
file=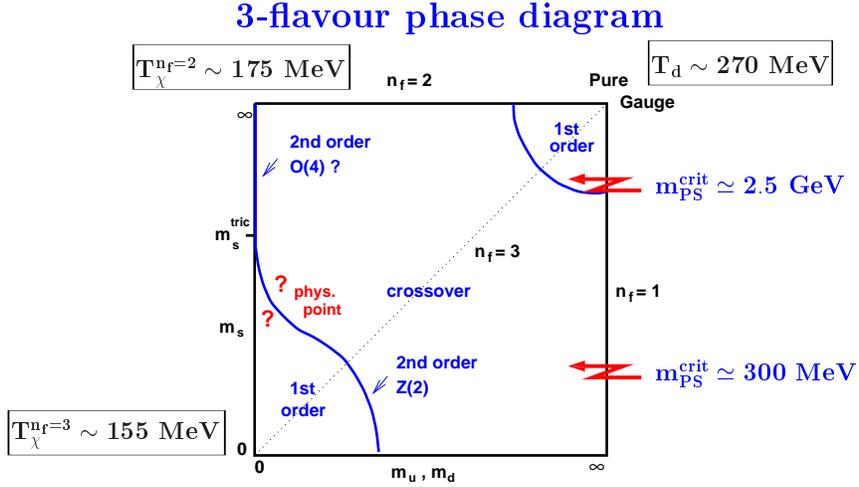,width=110mm}
\end{center}
%\vskip -0.9truecm
\caption{The QCD phase diagram of 3-flavour QCD with degenerate
(u,d)-quark masses and a strange quark mass $m_s$~\cite{karschlect}.}
%\vskip -0.8truecm
\label{fig:phased}
\end{figure}
%%%%%%%%%%%%%%%%%%%%%%%%%%%%%%%%%%%%%%%%%%%%%%%%%%%%%%%%%%%%%%%%%%%%%%%%%
\subsection{The phase diagram of QCD}

 Based on what we learnt above, fig.(\ref{fig:phased}) indicates schematically where one can expect first and higher order transitions. What varies in the diagram is the value of the $m_u=m_d$ mass and the mass of the strange quark $m_s$. The upper right corner contains the $Z(3)$ transition. As discussed before, it is known to be first order~\cite{kuti}. The deconfining transition $T_d$ is rather high.   The lower left corner contains the case of the $N_f=3$ chiral limit, which is first order as well, according to the renormalization group analysis of
 eq.(\ref{eq:wilczek}). The transition temperature $T^{N_f=3}_{\chi}$ is lowest.  The case $N_f=2$, upper left corner, is second order~\cite{wilczek}. Its transition temperature is not as low as that of three flavours. So the borderline between first order and crossover  
ends in a second order point at $m_u=m_d=0,m_s=\infty$. Gavin et al.~\cite{gavin} find the critical behaviour of  the lower left borderline is governed by an effective action with a $Z(2)$ symmetry. This $Z(2)$ symmetry is not present in the original QCD action.  

Clearly the determination of the exact location of this line vis \`a vis the physical values of the quark masses is of paramount importance.

\subsection{Chiral and Z(3) order parameters in flavoured QCD}

\begin{figure}[t]
\begin{center}
\epsfig{bbllx=105,bblly=220,bburx=460,bbury=595,clip=,
file=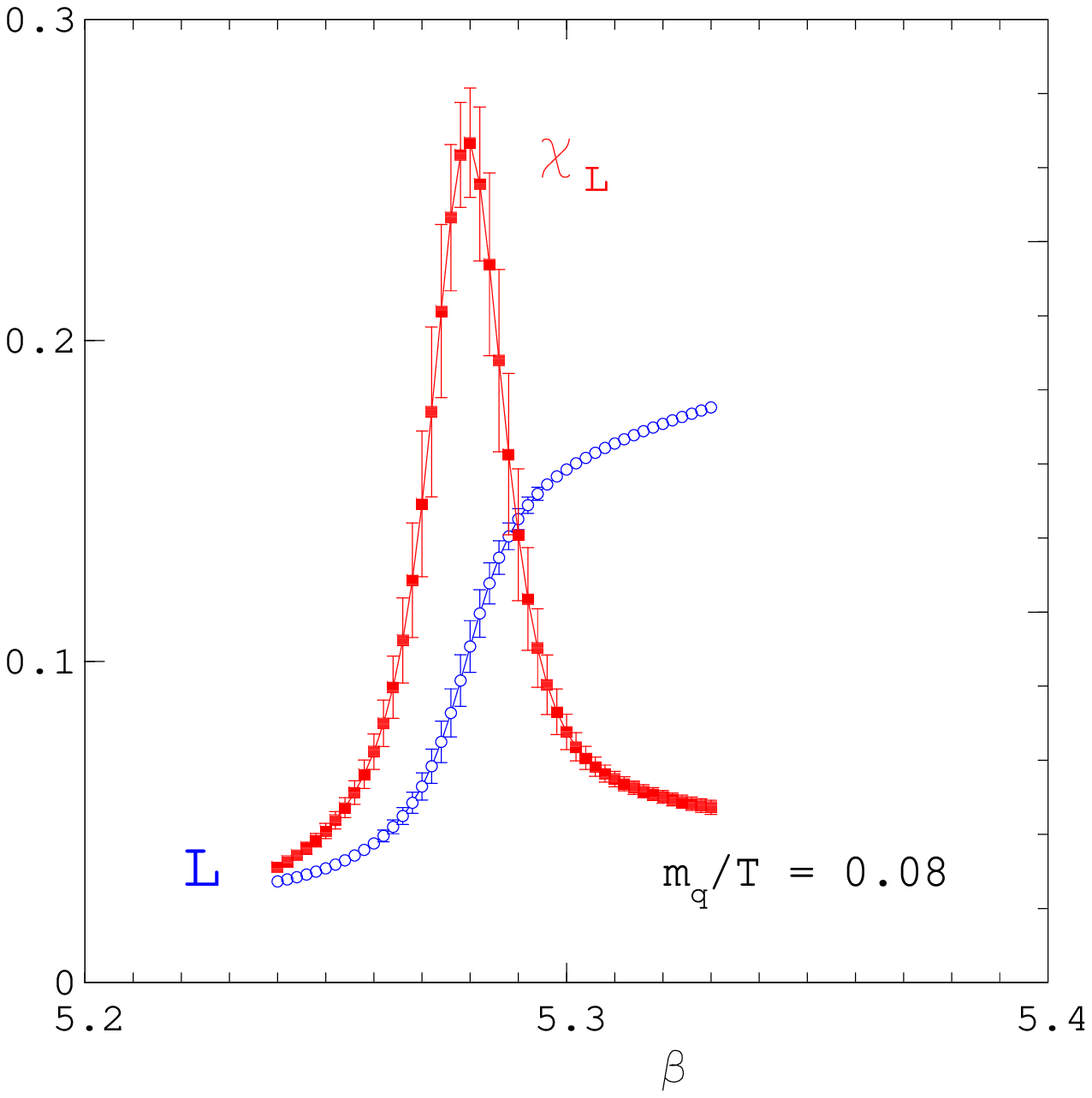,height=60mm}
\epsfig{bbllx=105,bblly=220,bburx=460,bbury=595,clip=,
file=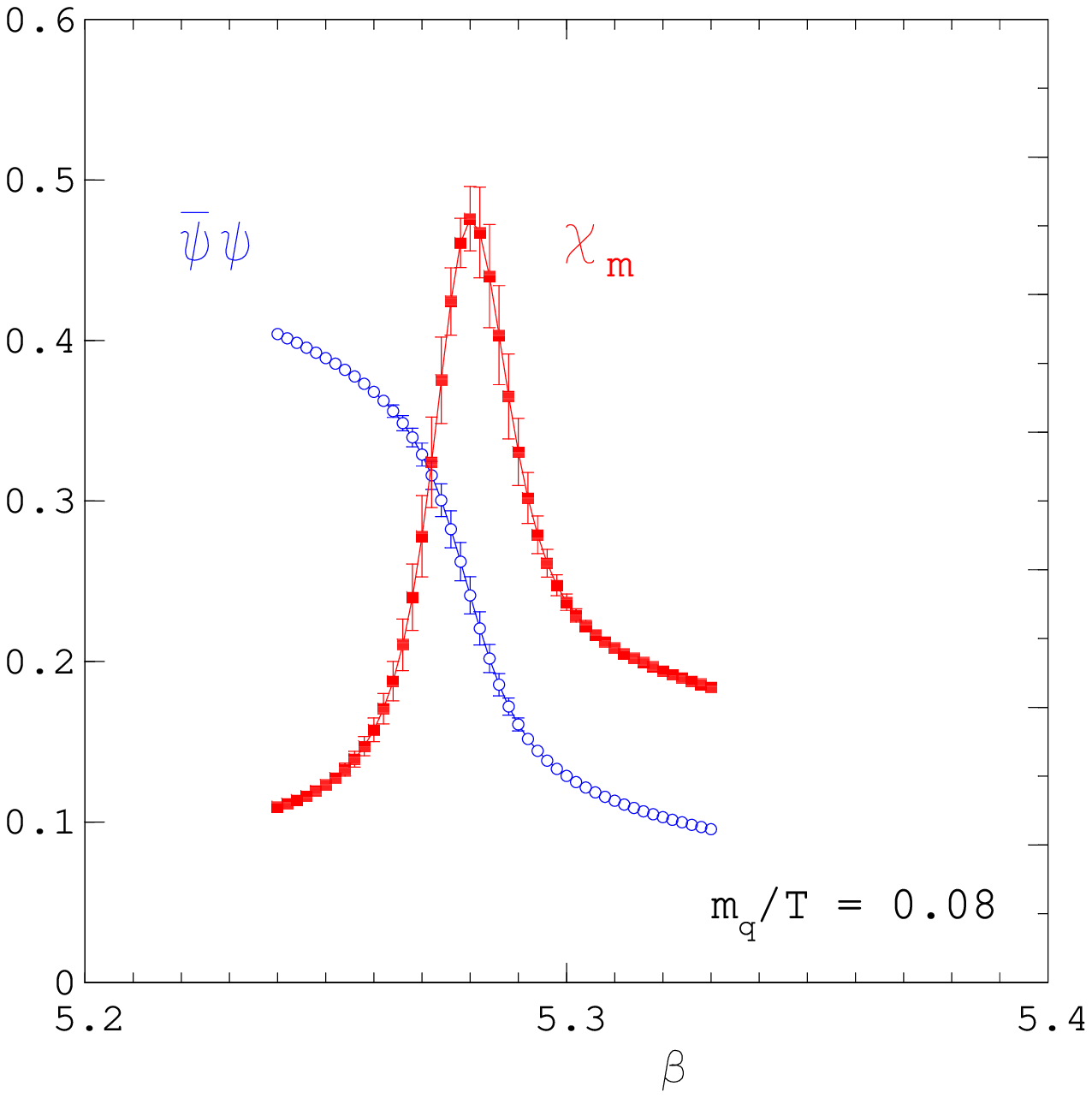,height=60mm}
\end{center}
\caption{Deconfinement and chiral symmetry restoration in 2-flavour QCD:
Shown is $\langle P\rangle$ (left), which is the order parameter for 
deconfinement in the pure gauge limit ($m_i\rightarrow \infty$), and 
$\langle \bar{\psi}\psi \rangle$ (right), which is the order parameter
for chiral symmetry breaking in the chiral limit ($m_i\rightarrow 0$).  
Also shown are the corresponding susceptibilities as a function of the 
lattice coupling $6/g^2$~\cite{karschlect}.}
\label{fig:demo}
\end{figure}

In fig.(\ref{fig:demo}) the transition region for two flavour QCD is shown.
We have only one symmetry, chiral symmetry. So we expect one transition at $T=T_{\chi}$, where the chiral order parameter  drops to zero .
There are two striking observations:
\null
\begin{itemize}
\item
 Despite explicit breaking of $Z(3)$ invariance the Wilson line drops steeply below some $T=T_d$.
\item
 The transition temperature is  the same for both order parameters: $T_c=T_d=T_{\chi}$ as peaks of the susceptibilities show.
\end{itemize}

\null

The first point is in seeming contradiction with the expectation that a heavy test quark forms easily a bound state with a dynamical light quark.
 However, you can argue that in the broken chiral phase the dynamical quarks acquire a mass heavy enough to recover approximate $Z(3)$ symmetry. If so, the Wilson line is a sensible order parameter in the broken chiral phase.

Confinement implies chiral symmetry breaking. After all, confinement implies a bound state of two massless quarks. But in a bound state the quarks must be able to flip their helicity. If so, then chiral symmetry cannot be restored before deconfinement sets in:   $T_d\le T_{\chi}$. In between, the Wilson line {\it could} be almost unity with the
chiral symmetry still broken! However,  Nature tells us that the quark condensate gets unstable above $T_d$, and that $T_d=T_{\chi}$. Why is not understood.

 In sharp contrast, if quarks are in the adjoint representation~\cite{karschadj} the  system has two exact symmetries, $Z(3)$ and chiral symmetry. So two different transition temperatures are expected. Flux tubes cannot end on adjoint matter, so form glueballs. The region in between has no glueballs anymore, but still a fermion condensate and a hadron spectrum. The adjoint fermion condensate stays stable till $T_{\chi}\sim 8 T_d$~\cite{karschadj}.

%%%%%%%%%%%%%%%%%%%%%%%%%%%%%figure%%%%%%%%%%%%%%%%%%%%%%%%%%%%%%%%%%%%%%%
%%%%%%%%%%%%%%%%%%%%%%%%%%%%%%%%%%%%%%%%%%%%%%%%%%%%%%%%%%%%%%%%%%%%%%%%%%

\section{Canonical Z(N) symmetries in SU(N) gauge theory}\label{sec:canoniczn}

In this section we start with QCD in 3+1 dimensions. We then
render  one  of the {\it{spatial}} directions periodic and study the effects due to varying  its size. The results will be of later use, especially in section (\ref{sec:work}).  Then we will switch on the temperature and see what happens. 

First we fix some general notions.

\subsection{Electric and magnetic fluxes}\label{subsec:prolegozn}

Below we give a quick review of  electric and magnetic fluxes and their 
free energy. Let only fields with trivial N-allity couple to the gauge fields.

\subsubsection{Electric fluxes}

For example, take the y-direction periodic mod $L_y$ and variable. The periodic  $\tau$ direction 
is supposed to be very long $L_{\tau}={1\over T}\rightarrow\infty $, as well as $x$ and $z$ directions.

To explain the essentials we will first put $L_x=L_z=0$ and consider the two-dimensional system.

 Then a time independent gauge transformation $\Omega_k$ is allowed to be discontinuous mod $z_k$ as in eq.(\ref{eq:k}) but now in the periodic y-direction:
\begin{equation}
\Omega_k(y+L_y)=z_k^*\Omega_k(y), k=0,1,..,N-1.
\end{equation}

An example of such a transformation is:
\begin{equation} 
\Omega_k=\exp{-i{y\over L_y}{2\pi\over N}Y_k}
\label{eq:discgauge}
\end{equation}

\noindent
with $Y_k=\mbox{diag}(k,....,k,k-N,....k-N)$, the diagonal k-hypercharge, an
NxN traceless matrix with $N-k$ entries k and k entries $N-k$. It has the property that 
\begin{equation} 
\exp{i{2\pi\over N}Y_k}=z_k
\end{equation}

\noindent is a center group element. It is a natural generalization of the 
familiar hypercharge $Y_1$. They span the  Cartan subalgebra, which consists of all
elements of the Lie algebra su(N) that can be diagonalized simultaneously.
It is N-1 dimensional. There is a lattice $L_c$ of elements in this subalgebra
that give upon exponentiation a center group element. So the elements $Y_k$
are special points on this lattice. They have an important property: let $q$ be a number between $0$ and $1$. Then the elements $Y_k(q)$ trace a rectilinear path in the
Cartan algebra on which only begin and end points correspond to centergroup elements (respectively $1$ and $z_k$).  Do this for all elements with $k$=1,....N-1.  Then we have an elementary cell of the lattice $L_c$.  The reader can see by inspection, that this true for N=3 and 4. This is the property that is useful for the $Z(N)$ invariant Wilson line distribution $E(\widetilde P)$ and will be used
throughout the dynamical calculations in subsection (\ref{subsec:thooftloop}).
The stability group of $Y_k$ is $SU(N-k)\times SU(k)\times U(1)$, the subgroup commuting with $Y_k$.

The gauge transformation $\Omega_k=\exp{i\omega_k}$ is represented in Hilbert space by $\exp{i\int d\vec x Tr\vec D\omega_k.\vec E}$. 
As  operator $\Omega_k$ commutes with all local gauge invariant operators, in particular the Hamiltonian. It does not commute with the Wilson line in the y-direction:
\begin{equation} 
\Omega_k P(A_y)\Omega_k^{\dagger}=z_k P(A_y)
\label{eq:wlinetransf}
\end{equation} 
It is important that  on the physical Hilbert space these operators have a unique effect, {\it{only}} depending on the discontinuity. 
To understand this take $\Omega_k$ and $\Omega_k^{\prime}$ with the same value for k. Form the product $\Omega_k^{\dagger}\Omega_k^{\prime}$. In this product the singular behaviour drops out: both transformations
belong to the same equivalence class through a regular gauge transformation. And a regular gauge transformation leaves a physical state invariant. The product of two elements from equivalence classes k and k' gives the equivalence class   k+k' mod N.  And finally $\Omega_k^N$ is regular.

As a consequence the eigenphases must be of the form $\exp{ik{2\pi\over N} e}$.
The number $e$ is integer and conserved mod N. 

And the physical Hilbert space divides into N orthogonal subspaces ${\cal{H}}_e$, e integer mod N,  on which $\Omega_k$ is diagonal with eigenvalue $\exp{ike{2\pi\over N}}$. The projector $P_e$ on such a subspace is given by
\begin{equation}
P_e|phys>={1\over N}\sum_k\exp{(-ike{2\pi\over N})}\Omega_k|phys\rangle.
\label{eq:projectorr}
\end{equation}

 And since the Wilson line $P(A_y)$ in the y-direction obeys $\Omega_kP(A_y)\Omega_k^{\dagger}=z_k P(A_y)$ a state with charge e can be written as a state with $e=0$ and with the line $P(A_y)$ wrapping $e$ mod $N$ times around the y-direction.
So $e$ is the promised conserved charge, and counts the number of ``strings'' or Wilson lines wrapping
around the y-direction (mod $N$). There is a  free energy $F_e$ for each of these electric flux sectors, defined by tracing only over the physical states of a given sector:

\begin{equation}  
\exp{-L_{\tau}F_e}=Tr_e\exp{-L_{\tau}H}.
\label{eq:fluxsectors}
\end{equation}

These free energies can be inferred  from simulations  on the lattice. First we need a formula relating the $F_e$ to  partition sums $Z^{(e)}_k$~\cite{thooft} . To this end substitute eq.(\ref{eq:projectorr}) into eq.(\ref{eq:fluxsectors}) and rewrite the latter as:

\begin{equation}  
\exp{-L_{\tau}F_e}={1\over N}\sum_{k=0}^{N-1}Z^{(e)}_k\exp{-ike{2\pi\over N}}.
\label{eq:twistedpartition}
\end{equation}

The partition functions on the r.h.s. of this equation are now Gibbs sums over physical states, with the operators $\Omega_k$ acting:
\begin{equation}  
Z^{(e)}_k=Tr_{phys}\Omega_k\exp{-H/T}
\label{eq:twistedconf}
\end{equation}

To understand the physical meaning of the partition functions there is an alternative definition of the operator $\Omega_k$. It is {\it only} valid on the physical subspace, where it reads:
\begin{equation}
\Omega_k=\exp{i{4\pi\over{gN}}TrE_y(y_o)Y_k}.
\label{eq:dipolepointsheet}
\end{equation}

Only on the physical subspace the two are identical. In fact they differ
by a regular gauge transformation, as you can infer from the exercise below.
\begin{itemize}

\item
Show that the operator (\ref{eq:dipolepointsheet}) has the same effect on the Wilson line $P(A_y)$ as $\Omega_k$ in eq.(\ref{eq:wlinetransf}). 
\item
Show that any regular gauge transformation $\Omega$ of $\Omega_k$ in (\ref{eq:dipolepointsheet}), $\Omega\Omega_k\Omega^{\dagger}$, has the same effect on $P(A_y)$. This means that physical states stay physical states after applying $\Omega_k$. Hint: make use of the fact that the centergroup element commutes with all of the gauge group.

\end{itemize}

So the discontinuous gauge transformation is brought about by a single dipole of strength $Y_k$ at the point $y=y_0$.

\begin{figure}
\begin{center}
\includegraphics{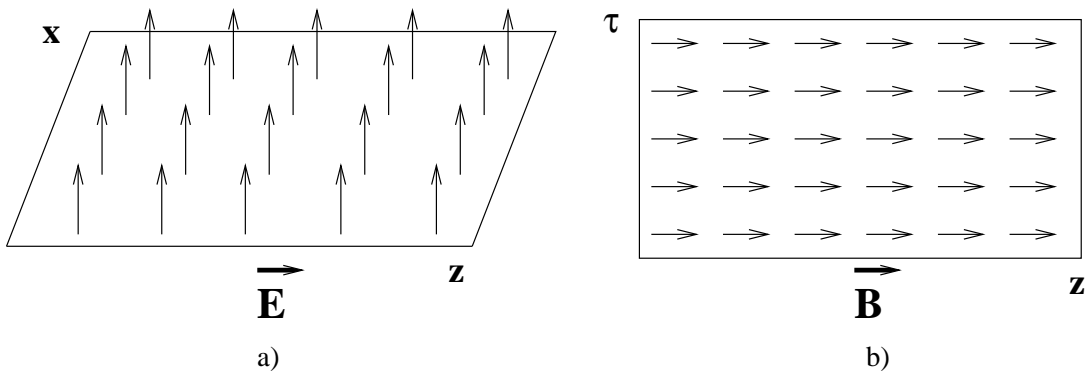}
\caption{Electric twist a) and magnetic twist b) partition functions.
The electric twist is a dipole sheet at fixed $(\tau,y)$. The magnetic twist 
is a Dirac flux in the z-direction, propagating in time, at fixed (x,y). In lattice language the plaquettes $P_{\tau, y}$ are twisted in a), $P_{x,y}$ in b).  }
\label{fig:dipolesheet}
\end{center}
\end{figure}

This generalizes easily from d=2 to d=4, adding the x and z dimensions.
The single dipole at $y_0$ becomes a dipole sheet on the (x,z) surface at the point $(y_0,\tau=0)$
as representing the operator $\Omega_{k_y}$ (see fig.(\ref{fig:dipolesheet})). We have added the suffix $y$
on $k$ to distinguish it from  a similar operator in z or x direction.
Once this is done we have to admit not only operators $\Omega_{\vec k}$
labeled by the vector $\vec k=(k_x,k_y,k_z)$, each component running from 0 to $N$. Obviously, we also  have fluxes $\vec e=(e_x,e_y,e_z)$ and the connexion between the free energy $F_{\vec e}$ and the partition functions $Z^{(e)}_{\vec k}$ generalizes to:
\begin{equation}
\exp{-L_{\tau}F_{\vec e}}={1\over{N^3}}\sum_{\vec k}Z^{(e)}_{\vec k} \exp{-i(\vec k.\vec e {2\pi\over N})}
\label{eq:dipolesfreeenergy}
\end{equation}
And so we have now a physical interpretation of the partition function
eq. (\ref{eq:twistedconf}) as the thermal average of the electric dipole sheet.
 It monitors the electric flux activity in the system as we will see in sections (\ref{sec:work}) and (\ref{sec:quasihooft}).
%In the thermal average we can deform the surface of the sheet.

%The $Z^{(e)}_k$ are written as pathintegrals  with an ``electric  twist'':
%they represent the overlap between the  physical state $\Omega_k|phys\rangle$
%and the physical state $\exp{-H/T}|phys\rangle$. The presence of $\Omega_k$ at $\tau=0$ creates vorticity in all the configurations $A_{(k)}$ in all  $(y,\tau)$ planes labelled by $(x,z)$. This vorticity is measured to be $z_k$ by any Wilson loop in the $(y,\tau)$ plane: $W(A_{(k)})=z_kW(A)$ where $A$ is any configuration contributing to the untwisted partition function. 

%The physical meaning of the electric twist is much more transparent in an equivalent formulation, ideally suited for  the lattice formulation.

Note that the partition functions with an electric twist are related through a Fourier transform to the free energies $F_{\vec e}^{(e)}$.  They do not define by themselves a free energy as they are off-diagonal matrix elements.

\subsubsection{Magnetic fluxes}\label{subsubsec:magflux}
Of course, one can define partition functions from physical states with a magnetic vortex line running
along a space direction, say the z direction. In continuum language the operator creating such a vortex is:
\begin{equation}
V_m=\exp{i\vec D(A)v_m(x,y)\vec E}
\label{eq:magnvortex}
\end{equation}
\noindent with $v_m(x,y)={\arctan({y\over x})\over N}Y_m$. When encircling the  point $x=y=0$ the  gauge transformation $\exp{iv_k(x,y)}$ picks up a factor
$\exp{i{2\pi\over N}Y_m}=z_m$. This gauge transformation remains, by definition, unchanged along the z-direction. We say that $V_m$ creates a vortex or ``Z(N) Dirac string''.      That means, a Wilsonloop $W$ in the fundamental representation that encircles the vortex will pick up the $z_m$ factor:
\begin{equation}
V_mWV^{\dagger}_m=z_mW.
\label{eq:wilsonlooptwist}
\end{equation}

 Any Wilsonloop with non-zero N-allity
$l$ will pick up a factor $(z_m)^l$. But Z(N) neutral loops will not sense 
the Z(N) Dirac string, hence the name. 

Next we address the question how to propagate the string in the Euclidean time. As a warm up we start with a small time lapse $\Delta\tau$ from $\tau$ to $\tau + \Delta\tau$. Then we have for the thermal average:
\begin{equation}
Tr_{phys}\exp{-H\tau}V_m\exp{-H\Delta\tau}V_m^{\dagger}\exp{-H({1\over T}-\tau)}.
\end{equation}

The question is now: what happens to the Hamiltonian:
\begin{equation}
H={1\over 2}\int d\vec x\big(\vec E^2+\vec B^2\big). 
\end{equation}

The electric field strength is in the adjoint representation so so  does not
feel the Z(N) discontinuity. One might be tempted to say the same of the mag  netic term. However, on the lattice the magnetic term is regulated as a Wilson loop on a plaquette. So all the plaquettes encircling the vortex pick up the Z(N) phase, according to eq.(\ref{eq:wilsonlooptwist}). As the vortex runs in the z-direction the $(x,y)$ plaquettes on the vortex are ``twisted''.

We are interested in finite time slices.
The string of twisted plaquettes in the Hamiltonian is then repeated in every $\tau$ slice, tracing out its history.

The notation for the magnetic partition functions is $Z^{(m)}_m$, with the time slice being the full period ${1\over T}$:
\begin{equation}
Z^{(m)}_{m}=Tr_{phys}V_m\exp{(-H/T)}V_m^{\dagger}\equiv\int DA\exp{-S_{m}(A)}.
\label{eq:magneticfluxfree}
\end{equation}

 They define directly a magnetic free energy $Z^{(m)}_m=exp{-L_{\tau}F^{(m)}_m}$, being diagonal matrix elements of the Hamiltonian. Note the difference with the electric twist partition function $Z^{(e)}_k$ in 
 eq.(\ref{eq:twistedconf}).  

The action $S_m$ is gotten from the twisted Hamiltonian. So at a given time slice we have along the vortex line:

\begin{equation}  
1-{1\over N}TrU(P_{x,y})\rightarrow 1-z_m({1\over N}TrU(P_{x,y})\big).
\label{eq:twistplaq}
\end{equation}

\noindent and repeat this for all time slices.

 The situation is shown in fig.(\ref{fig:dipolesheet} b). The vortex creates a singular dipole field $\vec B$ along the z-direction in the $Y_k$ colour direction. 

Our definition of magnetic flux free energy has  a caveat: looking at 
eq.(\ref{eq:magneticfluxfree}) we put in a trace over all physical states.
If we want no electric flux, we should have projected on the corresponding subspace.  The reason we did not have to do this is the additivity of electric and magnetic flux free energy:
\begin{equation}  
F_{\vec e;\vec m}=F^{(e)}_{\vec e}+F^{(m)}_{\vec m}
\label{eq:additivity}
\end{equation}

Additivity is supposed to be true in the thermodynamic limit $L_{x,y,z}\rightarrow\infty$.

 Using in our definition eq.(\ref{eq:magneticfluxfree}) the inverse of eq.(\ref{eq:dipolesfreeenergy}) we find:
\begin{equation}
Z^{(m)}_{\vec m}=\sum_{\vec e}\exp{-L_{\tau}F_{\vec e;\vec m}}
\end{equation}

%Still we need to understand how to simulate numerically the partition functio%ns
%in the form of a path integral. For the magnetic partition function this is i%mmediate. The lattice Hamiltonian consists of plaquettes. Those that encircle %the vortex will, like the Wilson loop in eq. (\ref{eq:wilsonlooptwist}) pick %up a $z_k$ factor (or a $z_k^*$ factor, depending on the orientation of tpla%quette). So the magnetic twist partition function is simulated by replacing t%he 
%usual lattice action by the  lattice action with a twist. Replace along the s%tring:

For the electric twist partition function one just exchanges $x$ and $\tau$.
As the plaquettes $P_{y,\tau}$ deliver the electric field in the continuum
limit, it is intuitively clear that this prescription will give the thermal average of the electric dipole sheet.

\subsubsection{Behaviour of flux free energies in the confined phase}
The behaviour of the electric and magnetic free energies is quite different
in the confined phase, where all sizes are macroscopic.

When the size $L_y$ of the periodic direction is macroscopic, the VEV of the Wilson line in the y-direction is zero. The system is confining with string tension $\sigma$. This means that $F_e-F_0=\sigma_e L_y$ and 
states with $e=0$ mod N are energetically favoured. The  energy $E_e=\lim_{L_{\tau}\rightarrow\infty}F_e$ of a state with $e=0$  mod N is the lowest, all others are higher by an amount $\sim \sigma L_z$, and the symmetry is ``restored'' because we have one unique ground state. Only the space ${\cal{H}}_{e=0}$ is of importance for confining physics. It contains the glueball states, the localized eigenstates of the Hamiltonian.
 The periodicity in e mod N comes about because N strings decay into glueballs.

The magnetic free energy is decaying exponentially fast~\cite{thooft}:
 $$F_{m_z}^{(m)}-F_0\sim \sigma L_z\exp{-\sigma L_xL_y}.$$
 This means the magnetic flux is screened. We will come back to this type of screening in the next section.

\subsubsection{A simple property of electric and magnetic twisted partition functions}
On the other hand the partition functions have a simple property. Suppose only one size  $L$ becomes small (meaning of hadronic size or smaller), whereas the others stay macroscopic.  Consider any partition function $Z_k$ with one single twist of strength $k$ like in fig.(\ref{fig:dipolesheet}a) or b)). If $L$  corresponds to one of the directions orthogonal to the planes shown ( y or $\tau$ in a), x or y in b)) then $Z_k$ obeys an area law with the area as shown in the figure, and a {\it{universal}} coefficient $\rho_k$:
\begin{equation}
Z_k\sim\exp{(-\rho_k Area)}.
\label{eq:initialrho}
\end{equation}
In the lattice formulation of the twist this universality is just a consequence of the Euclidian invariance under exchange of the $\tau$ and $x$ axis of fig. (\ref{fig:dipolesheet}) a) into b) and vice versa. The function $\rho_k$ is computed perturbatively  
in subsection (\ref{subsec:thooftloop}) in terms of the running coupling $g(1/L)$.

\subsubsection{Behaviour of the partition functions in the hot phase}
Physically one would expect that electric flux free energies will show screening behaviour. And this is verified easily by using the simple property
of the partition function mentioned in the previous section at high temperature. All partition
functions with a twist in the time direction will behave according to eq. (\ref{eq:initialrho}). These are precisely the partition functions appearing in the Z(N) Fourier transform that leads to the electric flux free energies in eq. (\ref{eq:dipolefreeenergy}):
\begin{equation}
\exp{-L_{\tau}F^{(e)}_{\vec e}}={1\over{N^3}}\sum_{\vec k}Z^{(e)}_{\vec k} \exp{-i(\vec k.\vec e {2\pi\over N})}
\label{eq:dipolefreeenergy}
\end{equation}

With the partition functions on the right hand side decaying like $\exp{-(\rho_k Area)}$ one can easily infer that the free energy differences 
$F_{\vec e}-F_{\vec 0}$ decay exponentially as well.
\begin{itemize}
\item
Deduce that $F^{(E)}_{e_x,0,0}-F^{(E)}_{\vec 0}\sim \exp{-\rho_{e_x}L_xL_y}$. 
\end{itemize}

So the electric fluxes behave radically different from the confining phase. They become exponentially fast degenerate, and the electric Z(N) symmetry is sponateously broken. 

The magnetic flux energies behave qualitatively the same as in the low T phase: they are screened.

\subsection{Breaking canonical Z(N) symmetry}\label{subsec:brocan}

 What happens as $L_y$ becomes smaller? In fact, if $L_y$ becomes on the order of the hadron size $R$ we  have a transition just as we had before with
the inverse temperature and $L_y$ interchanged. So we expect the Wilson line $P(A_y)$ to acquire a VEV. But the physics will be that of 2+1 dimensional Yang-Mills plus the degrees of freedom of $A_y$ that do not depend on y anymore, i.e. an adjoint Higgs. In the 2+1 dimensional space, $A_y$ is a scalar with respect to rotations in the x-z plane.  But  there will still be a string tension $\sigma$.

 So what {\it{has}} changed
qualitatively in this phase?  
\begin{itemize}

\item
In the x-z plane  domain walls, or rather domain ''lines'',  appear: they separate two regions where the spatial Wilson line, curled up in the y-direction,  has different $Z(N)$ values by a factor $\exp{ik{2\pi\over N}}$.

 These walls have a tension $\rho_k(M)$ which is perturbatively calculable with the 4d running coupling $g(M)$ ( $M=1/L_y$ is any mass scale larger than the critical one $M_c$ ).  
\end{itemize}

The tension $\rho_k(M)$  is computed from
the normalized  twisted magnetic partition function $\hat Z^{(m)}_{k}=Z^{(m)}_{k}/Z^{(m)}_0$. At small enough
$L_y$ it behaves as $\hat Z^{(m)}_k=\exp{-\rho_k L_xL_{\tau}}$.  
This is the expression for a domainwall stretching in the x-direction, with energy density or tension $\rho_k$.

 What is obvious without calculation is that the tension of the wall is
\begin{equation}
 \rho_k=d_k (1/g(M))M^2 
\label{eq:rho}
\end{equation}
 from dimensional reasoning  and the fact that the calculation is semi-classical. Its width will be $O((g(M)M)^{-1})$. The calculation will be done in detail
in section (\ref{sec:work}).

So we find that the magnetic flux free energy in $x$ and $z$ direction is no longer screened! The domain lines are  made of unscreened magnetic flux.

This summarizes the effect of the breaking of  canonical $Z(N)$ symmetry.

\subsection{Intrinsic Z(N) symmetry in 2+1 dimensional Yang-Mills}\label{subsec:intrinsiczn}

In the 3d Yang-Mills system there is also an {\it{intrinsic}} canonical $Z(N)$ symmetry  as opposed to the ``extrinsic'' Z(N) symmetry discussed above. It was discovered long ago~\cite{thooft} and is due to the appearance of magnetic vortices in 2+1 Yang Mills theory. Hence the name  ``magnetic Z(N)'' symmetry with as order parameter the vorticity.

We give here a quick recapitulation of how this symmetry is realized and 
its relation with confinement in 2+1 dimensions~\cite{thooft}~\cite{alex}. The results are going to be useful in section (\ref{sec:work}).

 A vortex is created by a gauge transformation $V_k(x,z)\equiv\Omega_k(x,z)$ with a discontinuity $\exp{ik{2\pi\over N}}$ across a line starting from the point $(x,z)$). The discontinuity is not seen by the adjoint fields. $V_k(x,z)$ has a purely local effect. Only when we surround it by a Wilson loop in the fundamental representation it gives a phase factor to the loop:
\begin{equation}
V_k(x,z)W(C)V_k(x,z)^{\dagger}=\exp{ik{2\pi\over N}}W(C)
\label{eq:canonicthooft}
\end{equation}
if the point $(x,z)$ is inside the loop $C$.

So $V_k(x,z)$  creates excitations that have a charge mod N with respect to a Wilson loop that surrounds the whole 2d system. This large Wilson loop is the generator of this intrinsic Z(N) symmetry. The large Wilsonloop commutes with the Hamiltonian. As a consequence the vorticity is conserved mod N.

If this symmetry
is realized in the spontaneously broken mode, $<V_1>\neq 0$, then we have N equivalent ground states. Each of these ground states corresponds to a different orientation of the VEV and they are all mutually orthogonal. 

Pick a given ground state $|k>$ with $<k|V_1|k>=v~z_k$, $0< v\le 1$.

 Then the action of $W(C)$ transforms it into a state $W(C)|k>$ where inside the perimeter $C$ the VEV of $V_1$ corresponds to that in the state $|k+1>$ because of eq.(\ref{eq:canonicthooft}).
\begin{itemize}
\item
Show that $V_k$ commutes with $|W(C)|^2\equiv W^{\dagger}(C)W(C)$. Deduce that the unitarized Wilson loop $\widetilde W(C)\equiv |W(C)|^{-1}W(C)$
has the same commutation relation with $V_k$ as $W(C)$.
\item
Find $<k|{\widetilde W(C)}^{\dagger}V_1\widetilde W(C)|k>=z_1<k|V_1|k>$, if the vortex operator acts inside the contour $C$.
\end{itemize}

That is, the loop creates a domain ''wall'' ~\cite{thooft} between the two groundstates.

What is the typical energy density and width of this wall? In 3d Yang-Mills theory there is a dimensionful coupling $g_3$ with its square having dimension of mass.  So a dimensional argument leads to an energy density of $g_3^4$ and the width $1/g_3^2$.

The VEV of the loop, $<k|W(C)|k>$, consists of the overlap of the state 
 turned into $|k+1>$ in the inside of the loop. So if we make the loop larger and larger the orthogonality of the ground states tell us the VEV goes to zero. 
That it decreases as fast as the exponent of the area follows from closer inspection of the overlap~\cite{alex}.  The overlap consists of a product  of local overlaps $\gamma(\vec x)$ between the vortices in a given point $\vec x$ inside the loop:
\begin{equation}    
<k|W(C)|k>=\Pi_{{\vec x}\in S(C)} \gamma(\vec x)=exp{-\sigma S(C)}
\end{equation}
This argument is correct to the extent that the vortices interact only locally,
typically over a distance defined by the dimensionful coupling $g_3^2$.
The tension $\sigma \sim g_3^4$ again on dimensional grounds.

Recall that our 2+1 dimensional system is embedded in a 3+1
dimensional world where the z-direction is periodic and of size $M^{-1}$. 
The 2+1 gauge coupling expressed in terms of the 4d coupling is:
\begin{equation}
g_3^2=g^2(M)M. 
\end {equation}
So for large $M$ the gauge coupling $g(M)$ is small.
 
This means that the energy of the  domain walls due to breaking of intrinsic Z(N)  is parametrically a factor $g(M)^{5}$ smaller than the tension
due to the periodic z-direction in eq.(\ref{eq:rho})! And the width of the 
wall created by the Wilson loop is of the order of $(g(M)^2M)^{-1}>>M^{-1}$ i.e. much larger than the extra periodic dimension. From this one concludes that,
 once extrinsic Z(N) is broken, the system is essentially 2+1 dimensional.
This hierarchy of scales is consequential for the next subsection.

\subsection{ The fate of broken  Z(N) at high temperature}

\begin{figure}[tb]

\centerline{~~\begin{minipage}[c]{9.0cm}
    \psfig{file=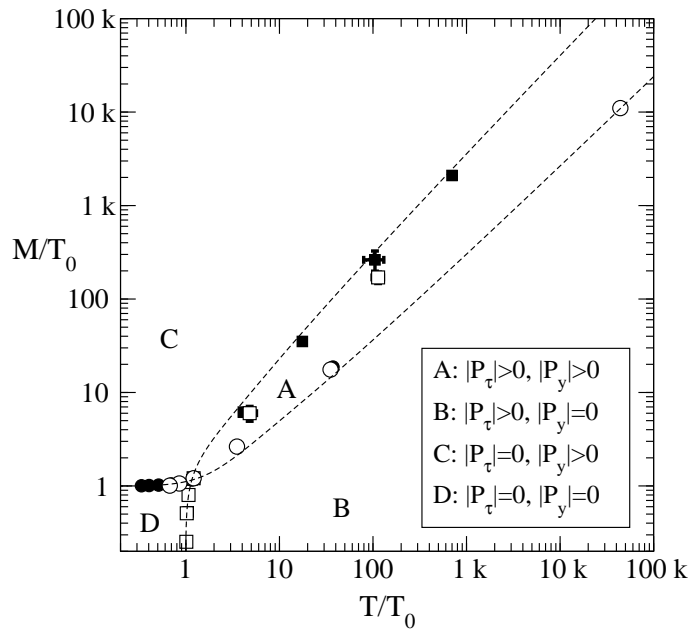,angle=0,width=9.0cm} \end{minipage}}

\vspace*{0.5cm}

\caption[a]{The 4d phase diagram in  units of $T_0\equiv T_c$, as
shown by Montecarlo data, ref.~\cite{farakos}. The data show the points where 
the Wilson lines $P_{\tau}$ and $P_y$ undergo the transition.
Note the symmetry $\tau \leftrightarrow y$. }

\label{fig:phys_4d}
\end{figure}

Let us now heat up this 2+1 dimensional  system.  For the case of two colours the procedure is illustrated in  fig.(\ref{fig:phys_4d}). It shows a simulation of  3+1 dimensional SU(2) gauge theory with $\tau$ and $y$ directions periodic and variable, of length $L_{\tau}=1/T$ and $L_y=1/M$ respectively~\cite{farakos}.

The 3+1 dimensional theory lies in region D. 
Simulated are the thermal  Wilson lines $P_{\tau}$ and the temperature 
$T_c(M)$  where it  becomes non-zero. The transition line $T_c(M)$ starts at the $M=0$ axis at $T=T_0$, rises steeply and bends to the right.  Also shown is the Wilson line $P_y$ with its critical behaviour.  The lattice data are shown as circles and the broken lines are for our purpose here just fits to the data. The two broken lines are mirrored through the diagonal. This should be obvious: the loci of the two types of transition cannot distinguish between $\tau$ and $y$: $P_{\tau}(T,M)=P_y(M,T)$.

Start from region C on the vertical axis  somewhere above $M=T_0=T_c$ in fig.(\ref{fig:phys_4d}). This portion of the vertical axis is the cold system in the broken ``extrinsic'' Z(N) phase. Here $P_y>0$ and also the VEV of the 't Hooft vortex operator $V_k$.
As we increase T along this line we first cross the broken line into region A where also the thermal Wilson line $P_{\tau}$  gets a VEV.  
 But the VEV  $<V_k>=0$, as was argued in ref.~\cite{kovner}~\cite{stephanov}: one can show with the arguments of subsection
(\ref{subsec:thooftloop}) that the correlation $<V_k(\vec x) V_k(\vec y)^{\dagger}>\sim
\exp{-m|\vec x-\vec y|}$ for large separation, as soon as the thermal Wilson line becomes non-zero. So this is the region where the intrinsic Z(N)
is restored.

 But the extrinsic Z(N) is still broken because of the difference in  energy scales. Note that the transition of the essentially 2+1 dimensional system is at a {\it{higher}} temperature than the 3+1  dimensional system. This is  
%%%%%%%%%%%%%%%%%%%%%%%%%%%%%%%%%%%%%%%%%%%%%%%%%%%%%%%%%%%%%%%%%%%%%%%%%%%%%%
%%%%%%%%%%%%%%%%%%%%%%%%%%%%%%%%%%%%%%%%%%%%%%%%%%%%%%%%%%%%%%%%%%%%%%%%%%%%%%%
intuitively reasonable. After all, in 1+1 dimensions the transition is at infinite temperature.

Finally increasing T even more,  the restoration of the extrinsic Z(N) takes place at the second crossing  of the horizontal line into phase B. Then we are in the high temperature phase B  of 3+1 dimensional QCD. 

The symmetry in the figure between phase B and C is deceptive from a physics point of view. In phase C there are one dimensional  ``domain'' walls tracing out a two dimensional sheet in time $\tau$. They do separate regions where the Wilson lines $P_y$ have different center group values and an observer in the (x,z) world 
can observe those walls with their high energy density. Phase B is the hot QCD phase where there are regions with different center group values for $P_{\tau}$, separated by two dimensional sheets as discussed at the end of subsection~(\ref{subsec:wilsonzndeconf}). These sheets cannot be interpreted as domain walls 
because they do not extend in time as solitons. In  section (\ref{sec:quasihooft}) their role will be seen to be that of detectors of electric flux.  The tension $\rho_k$ is indeed symmetric with respect to the diagonal.

The reader may ask the justified question: how do the currently popular extra periodic space dimensions of size $1/TeV$ to our four dimensional world fit into this picture? The answer is simple and expected from the one dimensional domain walls found above: for 5D SU(N) gauge theory, phase C now contains {\it{two-dimensional domain walls}}! The  energy density of these walls is typically $(TeV)^{4}$. The form of the phase diagram is qualitatively the same. Hence the extrinsic Z(N) symmetry gets restored at high T and the walls put constraints on cosmological models~\cite{extradim}~\cite{farakos}.  
 \nopagebreak

\section{Forces and screening in the plasma}\label{sec:forces}

Till now we did not mention what happens to the forces in the QCD plasma.

In this section we will first discuss Debye screening on a perturbative basis.
IN QCD this  turns out to be insufficient. We need a non-perturbative definition. Then, in the next subsection an operator formalism is presented, useful for the systematics of the lattice calculations. 

Finally magnetic screening is defined. This is a new and important aspect of 
thermal QCD.

\subsection{Electric screening}

In a QED plasma one would like to know what happens to the  electric field due to a heavy charge Q and to the Coulomb force  between two static charges of opposite signature at distance $r$ in the z-direction. Let us look at scalar QED,
as it shows in leading order in the coupling some features in common with QCD.

The Coulomb force is transmitted by the the $A_0$ potential. The propagator of 
$A_0$ is renormalized by the one loop scalar and seagull diagram and gives the self energy, as calculated by the Feynman rules discussed below eq.(\ref{eq:gibbs}):
\begin{equation}
\Pi_{\mu\nu}(p)=e^2T\sum_{l_0}\int{d\vec l\over{(2\pi)^3}}\big({(2l_{\mu}-p_{\mu})(2l_{\nu}-p_{\nu})\over {(l-p)^2l^2}}-2{\delta_{\mu\nu}\over{p^2}}\big)
\label{eq:oneloopselfenergy}
\end{equation}

The self energy is transverse, $p_{\mu}\Pi_{\mu\nu}(p)=0$. It has two independent tensors, that we choose to be $\Pi_{00}$ and $\Pi_{\mu\mu}$. For $T=0$ they are  proportional.

For the Coulomb force we are interested in 
 the static limit $p_0=0$. We resum all the self energy bubbles to get the propagator 
 and the static part of the $<A_0A_0>$ propagator becomes : 
\begin{equation}
{1\over{\vec p^2}}\rightarrow {1\over{\vec p^2+\Pi_{00}(p_0=0,\vec p)}} 
\end{equation}

To find the pole to lowest order in the coupling, we let  $\vec p\rightarrow 0$  and find $\Pi_{00}(0,\vec 0)=e^2T^2/3$.
We use dimensional regularization so that the $l_0=0$ contribution to $\Pi_{00}$ is $\sim\int d\vec l{1\over {\vec l^2}}=0$. Only hard modes proportional to $T$ inside the loop contribute to the pole mass!

In configuration space this leads to Coulomb screening:
\begin{equation}
{1\over r}\rightarrow {1\over r}\exp{-m_Dr}~\mbox{with}~m_D={e^2T^2 \over 3}.
\label{eq:screening}
\end{equation}
In scalar QED the self energy is gauge independent to all orders in perturbation theory. And so you can start to evaluate the corrections to the Debye mass
by computing the corrections to the pole location. Indeed, one can reformulate perturbation theory by adding the screening mass term to the action of scalar QED, and use the Feynman rules with the modified static propagator above:
\begin{equation}
S_{QED}=\{S_{QED}+m_D^2\int_{\vec x} A_0^2\} -m_D^2\int_{\vec x} A_0^2
\end{equation}
 To avoid double counting  you have to subtract the screening term as well, and use it as an insertion whenever you have a self energy subdiagram. This procedure 
leads to a well-defined perturbation series. However the powerlike infrared
divergencies are now cut-off at $m_D\sim eT$ so we can expect terms in the series like $e^4/m_D\sim e^3$, i.e. odd powers in the coupling. But otherwise the series can be computed to arbitrary order by taking the Debye screening into account.

The non-abelian case to lowest order is  qualitatively the same. The Debye mass changes only  by replacing $e^2\rightarrow g^2(N+N_f/2)$. The salient differences are:
\begin{itemize}
\item
In the non-Abelian case the self energy is not gauge independent. For the pole location one can argue that it is gauge independent. 
\item
More important,  already in next to leading order an infinity of diagrams in  the static sector contributes~\cite{linde}~\cite{rebhan}. 
\end{itemize}

This means that an observable is needed
to define the screening mass {\it{independently}}  of perturbation theory.

Indeed, there exists a natural definition in terms of the correlation of two
static charges in the fundamental representation. From the results of the previous sections and appendix A we can write this as the correlator of two Wilson lines:

\begin{equation}
<P(r)P(0)^{\dagger}>\equiv\exp{-{F_E(r)\over T}} 
\label{eq:polyakovcorrelator}
\end{equation}

 This can be simulated on the lattice by non-perturbative means. To this end one takes a  lattice periodic in all directions, and the Wilson lines separated over a distance r in the z-direction.

In the confining phase, below $T_c$, this correlator falls off at long distances
as $\exp{-{\sigma(T)\over T}r}$, due to the string tension $\sigma(T)\equiv V_T(r)/r$. 
Above $T_c$ the string tension gets screened by the Debye mass and the potential becomes:
\begin{equation}
F_E(r)=F_{E0}-{c_E\over r}\exp{-m_Dr}.
\label{eq:debyemassdef}
\end{equation}

The parameters in this free energy are only depending on $T$.

\subsection{An operator formalism as a bookkeeping device}

The path integral of the  spatial correlator can be read in an alternative way.
Consider the fictitious Yang-Mills Hamiltonian in the $(x,y,\tau)$ space, with its physical Hilbert space. This Hilbert space will contain eigenstates of the Hamiltonian which are different from the one in which we live. One space direction, $\tau$, is finite and periodic with period $1/T$. That means that the rotation group
in these three dimensions is reduced to $SO(2)$ times the discrete rotation group admitted by the periodic finite $\tau$ direction. The path integral reads in terms of this Hamiltonian $H$ and the said physical Hilbert space:
\begin{equation}
<P(r)P(0)^{\dagger}>=\big(Tr_{phys}\exp{-H\hat L_z}P\exp{-Hr}P^{\dagger}\exp{-H\hat L_z}\big)/\big(Tr_{phys}\exp{-H L_z}\big).
\label{eq:xytauhamilton}
\end{equation}
The Wilson lines P are now expressed in terms of the canonical operators $A_0$, $\hat L_z=(L_z-r)/2$, and $L_z>>r$. In the limit of $L_z\leftarrow\infty$ the correlation becomes a some of exponential decays:
\begin{equation}
<P(r)P(0)^{\dagger}>=\sum_n|\langle 0|P|n\rangle|^2\exp{-m_nr}.
\label{eq:massstates}
\end{equation}
The mass gap $m_n$ is the value of the energy compared to the groundstate energy, at zero momentum $p_x=p_y=p_{\tau}=0$.

We want an efficient bookkeeping system for the states excited by the Wilson line (and eventual other observables).

To this end we define in our fictitious Hilbert space a parity transform $P$, under which {\it only} the y-direction changes sign (and hence only $E_y$ and $A_y$). Remember the rotation group is reduced to $SO(2)$, so simultaneous flipping of $x$ and $y$ is a rotation. 

Charge conjugation is as usual: $A_{\mu}\rightarrow -A_{\mu}^t$.

Still another quantum number is $R$-parity:it changes $\tau$ into ${1\over T}-\tau$, and $A_0$ into $-A_0$\footnote{ The combination of time reversal T ($A_0\rightarrow A_0^t, A_i\rightarrow -A_i^t$) and charge conjugation $C$ in the usual Euclidean version of the theory gives an operation $R=TC$ that only flips the sign of $A_0$. Time reversal has no effect on the Wilsonline, bwcause it inverts the time ordering and at the same time transposes $A_0$ as a colour matrix. So the Wilson line is transposed as a matrix , but its trace stays the same. So $R$ has the same effect on the Wilson line as $C$ alone.}.

So the symmetry group is $SO(2)\times Z_2(R)\times Z_2(P)\times Z_2(C)$, hence states labelled by $J^{PC}_R$. The $SO(2)$ group is generated by $x\partial_y - y\partial_x$. So $PJP=-J$. Look at any eigenstate  $\vert j\rangle$ of $J$   with $j\neq 0$. Then the state $P\vert j\rangle$ has $J=-j$, so is orthogonal to $\vert j\rangle$. From these orthogonal states we can form the parity doublet, degenerate in energy:
\begin{equation}
\vert j;\pm\rangle=(1\pm P)\vert j\rangle.
\end{equation}

For spin zero states this argument fails. And indeed, lattice simulations reveal~\cite{teper3d} differences up to a factor two in spin zero parity conjugates.

Clearly our Wilson line operator excites spin zero  and positive parity states. Also, under $C$ and $R$:

\begin{equation}
P(A_0)\rightarrow P(A_0)^{\dagger}.
\end{equation}

An important caveat is due to Arnold and Yaffe~\cite{arnoldyaffe}:  the potential consists of {\it{two}} channels of exponential decays! One is governed by the correlation of $Im P$ and the other by that of $Re P$. The first is odd under charge conjugation ($A_0\rightarrow -A_0^t$), the second even. So they do not mix. The Debye mass corresponds to the odd channel,
as it should, according to our definition in terms of self-energy.

There is no difference between the two channels if ~ $<PP>=<P^{\dagger}P^{\dagger}>~=0$. And in the phase where Z(N) is unbroken it is not hard to see that
both are exponentially small with respect to the correlation $<P^{\dagger}P>$.
The exponent is controlled by the string tension $\sigma$.  In that case the lowest energy state with energy $E_0(T)$ has $\vec e=(e_x,e_y,e_{\tau})=0$. All states with $\vec e_{\tau}\neq o$ are exponentially suppressed by factors $\exp{-\sigma L_z}$, and the reader can convince himself by going through the exercise below that
 both correlators give the same area law in the hadron phase. This is intuitively expected 
from two like charges being unscreened: in a periodic volume their expectation value should be zero.

\begin{itemize}
\item
Show  $<PP>=Tr_{phys}e^{-(H-E_0)\hat L_z}Pe^{-(H-E_0)r}Pe^{-(H-E_0)\hat L_z}$, up to exponentially small terms, if  $Z(N)$ is unbroken. $\hat L_z=(L_z-r)/2$.
\item
~From the above, show that  $<PP>$ is a superposition of amplitudes involving the flux states
$|e_{\tau}>$ by using the projection operator eq.(\ref{eq:projectorr}).
\item
Find the suppression factors in front of $<e_{\tau}+2|Pe^{-(H-E_0)z} P|e_{\tau}>$ for all $e_{\tau}$ and their absence in front of $<0|P^{\dagger}e^{-(H-E_0)z}P|0>$. 
\end{itemize}
In the hot phase there is no reason the two channels are the same. Two like charges with screening can coexist in a periodic volume.

We can abstract the following conclusion from the above. The fictitious Hamiltonian has for $T$ below $T_c$ (don't forget that this means for $\hat H$ that  the spatial dimension in the $\tau$-direction gets large enough, wheras the temporal direction in the z-direction stays infinitely long)  a $Z(N)$ symmetry which is the canonical symmetry discussed in section (\ref{sec:canoniczn}). As long as the symmetry is unbroken (for $T\le T_c$) this Hamiltonian has winding states, labeled by the conserved quantum number $e_{\tau}$. The winding states have a tension $\sigma$. In the limit that $T=0$ we have the usual 3d Hamiltonian with the glueball mass spectrum and with ground state $\mid e_{\tau}=0\rangle$. All other winding states have infinite energy in this limit. For finite $T$ the winding states have finite energy. and are the 
groundstates of identical towers of states.

On the other hand, when $T\ge T_c$, we have  Z(N) realized in the broken mode. The $N$ winding states $\vert e_{\tau}\rangle$ have become the $N$ degenerate groundstates with, again.  each an identical mass spectrum. The Debye mass is one of those mass levels, as we will see in the next section.

\subsection{Screening of heavy magnetic charges}
Not only the force law between heavy electric charges like the 
heavy quark, but also the force between heavy magnetic charges tells us about the medium.  The original idea of 't Hooft and Mandelstam~\cite{tH76} was that of a dual superconductor, with the electric Cooper pairs replaced by some form of magnetic condensate. Especially the lattice community has been fascinated
through the last 25 years with this idea because it defies perturbative access.

%If only matter in adjoint representations is coupled to the gauge fields, 't %Hooft~\cite{thooft} taught us how to find local operators that create monopol%es with a flux $\exp{i{2\pi\over N}}$ or multiples thereof. 
In section (\ref{subsubsec:magflux}) we constructed an operator $V_k$ creating
a magnetic flux of strength $\exp{ik{2\pi\over N}}$, eq.(\ref{eq:magnvortex}).

To get the  monopole anti-monopole pair at points $(0,r)$ we have the vortex end at $0$ and  $r$ on the positive z-axis. The vortex is given by a gauge transformation $V_k(\vec x)$ which is discontinuous modulo a center group element $\exp{ik{2\pi\over N}}$ when going around the vortex. The vortex is like the Dirac string in QED.  It is unobservable by scattering with
particles in the adjoint representation, as long as it has center group strength.

% You can put it as you want as long as it ends  on the  same two  end points.% This is easy to understand. If the vortex is in a given position, one can de%form it by applying a gauge transform which has the same  discontinuity on th%e closed loop $\Delta S$ that is the deformation of the string $S$ into $S'$.% The state you get is no longer the same. But after summing, as we are told t%o do in the Gibbs trace, the result with the deformed string $S'$  will be th%e same as that with the string S.

The Gibbs trace can be worked into a path integral along the same lines as in section (\ref{subsubsec:magflux}), and on the lattice it takes the form~\cite{groeneveld}:
\begin{equation}
\exp{-{F_M(r)/T}}=\bigg(\int DA\exp{-S_{(k)}(A)\bigg)}/\int DA\exp{-S(A)}.
\label{eq:latticemagnetic}
\end{equation}

\begin{figure}
\begin{center}
\epsfig{file=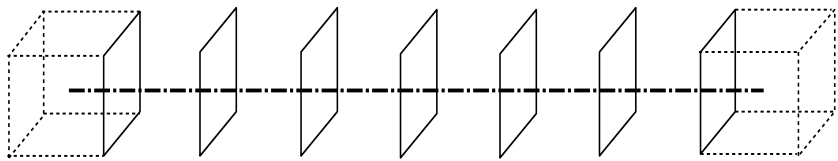,width=11cm,height=2cm}
\end{center}
\caption{\label{fig:chain}Monopole antimonopole pair induced by
twisting the plaquettes pierced by the Dirac string.}
\end{figure}

The action $S_{(k)}$ is the usual action, except for those plaquettes pierced by the Dirac string. Those plaquettes are multiplied by a factor $\exp{ik{2\pi\over N}}$, as in fig.(\ref{fig:chain}). 
\begin{itemize}
\item
Show that any deformation of the 
string can be obtained by  a change of integration (link) variable through a centergroup element.
\end{itemize}

 The reader will recognize from section (\ref{sec:canoniczn}) the magnetic vortex but now with endpoints, where the monopole pair resides. Varying the endpoints permits one to find the potential for all separations. 

Screening is expected  in both confined and deconfined phases:
\begin{equation}
F_M(r)=F_{M0}-c_M{\exp{-m_Mr}\over r}.
\end{equation}
All parameters are function of T.
 In the cold phase the screening is a consequence of the electric flux confinement. This is natural because the ground state contains a condensate of ``magnetic Cooper pairs'', according to the dual superconductor analogy. It is a screening mechanism whose details are not understood\footnote{ Nevertheless a quantitative understanding of the energy of a magnetic flux exists~\cite{thooft} as mentioned in subsection (\ref{subsec:prolegozn}).}. We dropped for notational reason the dependence on the strength  k of the monopole. It is important to note that this strength comes in periodic modulo N!
So the screening length is a periodic function of k and, because of charge conjugation, of $N-k$. It would be interesting to see what this dependence is.

In the hot phase there are indications from spatial Wilson loop simulations  that there is additional thermal screening from magnetic quasi-particles, as discussed in section (\ref{sec:quasihooft}).

Analogous to  the Wilson line correlator we consider the Hamiltonian $\hat H$ in the fictious system of $(x,y,\tau)$ variables. We search  the  operator $V_k$ acting on the
Hilbert space of physical states  of this Hamiltonian, that reproduces the path integral  eq.(\ref{eq:latticemagnetic})~\footnote{We use the same notation 
as for the vortex operator in $(x,y,z)$ space as there is no risk for confusion.} . So $V_k$ should create a vortex in the (x,y) plane at 
every time slice $\tau$ and the Hamiltonian $\hat H$  should propagate every one of these vortices in the z-direction over a distance r. So $V_k$ is the 't Hooft vortex operator discussed in section (\ref{sec:canoniczn}):

\begin{equation}
V_k=\exp{i\int_{x,y,\tau}Tr\vec D(A)v_k(x,y)\vec E}.
\end{equation}

\noindent with $v_k(x,y)=arctan({y\over x}{1\over N}Y_k$.

Both under parity and charge conjugation the vortex $V_k$ transforms into $V_k^{\dagger}$.  Its spin $J$ equals $0$, despite the appearance of the rotated 
singularity line. On physical states the location of the singularity does not matter.  Hence the operator  $Im V_k$ excites spin zero states with $P=C=-1$. The magnetic screening mass should correspond to the self-energy of a magnetic gluon, just like the correlator of the thermal Wilson line had to correspond to the self energy of a temporal gluon, So we choose the negative charge conjugation component  $Im V_k$.  The magnetic screening mass distinguishes itself from the electric screening mass by the opposite parity. This will prove important!

For $SU(2)$ gauge theory $Im V_1=0$ and  we have to take $Re V_1$  exciting $J^{PC}=0^{++}$ states (All of the physical Hilbert space is $C=+1$.).  
 
\begin{itemize}
\item 
Convince yourself of the quantum number assignments (parity in two dimensions
flips only one direction).
Prove that $R=TC$ parity is +1.
 \end{itemize}
Perturbation theory is not reliable  and we need lattice  
simulations~\cite{rebbi}~\cite{defor}.

\section{A quantitative method: perturbation theory and dimensional reduction}\label{sec:reduction}

Lattice simulations are our only tool today for tackling the critical region of QCD in a quantitative fashion, as far as the problematic fermions with small (including realistic) masses are avoided. But the region above  a few times the critical temperature can be accessed by the method of dimensional reduction~\cite{pis} without that problem. As we will see, fermions come in through the parameters of the reduced theory.

The method of dimensional reduction permits one to do perturbation theory, not only at very high temperatures but down to $T\sim2T_c$.
To obtain all coefficients of the perturbation series, one has to do dimensionally reduced
lattice simulations, i.e. simulations in three dimensions. This is due to the
three dimensional magnetic sector of the theory being a confining theory.

In fact the idea is similar to that in Kaluza-Klein theories: at high temperature the periodic dimension is very small with respect to the typical mass scale
of 3D Yang-Mills, and the Fourier modes in the periodic direction
are proportional to $2\pi T n$ with $n$ integer:
\begin{equation}
p_0\rightarrow 2\pi nT,~\mbox{n integer for bosons.}
\end{equation}
 For the fermions the Fourier modes are anti-periodic so the $n$ are half-integer.
 All non-zero modes are called hard modes.

So one works in three dimensional space with a 3D action. The field variables
are the constant modes $A_{\mu}(\vec x)$. The parameters in that action  take the effects of the temperature into account.  Some of the constant modes, the electric ones, get a
mass due to Debye screening. The magnetic modes stay massless, at least in perturbation theory. So if one is not interested in distances on the order of the Debye screening but in much longer distances, integration of the electric screening modes is mandatory.
We are then left with a 3D theory with only magnetic modes. They interact with a dimensionful coupling $g_M^2$ and describe a theory which is accurate on mass scales equal to or
smaller than $g^2_M$. 

\subsection{Integrating out the hard modes} 

To be precise we want to integrate out all degrees of freedom in the original QCD action that relate to momenta and Fourier modes of order T. So we need to fix a cut-off $\Lambda_E$ somewhere in between the scale $T$ and $gT$. In low order we can do without. This is because we are interested in amplitudes with external legs with $n=0$ and $p\sim gT$. To one loop order all modes with $n\neq 0$ in the loop introduce 
a scale of order $T$ in the loop integration over $\vec p$. The mode with $n=0$
has momenta of order $gT$ injected from the external legs, so the momentum integration will involve only $gT$~\footnote{To two loop order there can be internal propagators with $n=0$ and momenta on the order of $T$. To compute to this accuracy one either introduces the cut-off we just discussed~\cite{braaten} or one  exploits
$E(\tilde P)$ as  generating functional~\cite{kajeff}~\cite{bronoff}
for the electrostatic action.}.

The form of our effective action $S_E$ is dictated by all symmetries, global and local of the original QCD action and which are respected by the integration process. That implies all the symmetries we knew already,  except that the electric term in the static action will have no $\partial_0\vec A$ term. So $A_0$ appears as an adjoint Higgs term in our 3D gauge theory. The electrostatic QCD action density reads:
  
\begin{eqnarray}
{\cal{L}}_{E} & = & Tr(\vec D(A)A_0)^2+m_E^2TrA_0^2+\lambda_E(Tr(A_0^2))^2+ {} \nonumber\\
              & + &  \bar\lambda_E \big(Tr(A_0)^4-{1\over 2}(TrA_0^2)^2\big)+Tr F_{ij}^2+\delta{\cal{L}}_E.
\label{eq:estat} 
\end{eqnarray}

Because of R- conjugation invariance ($A_{0}\rightarrow -A_{0}$) the electrostatic action must be even in $A_0$. 

So far for the form of the action. The parameters in the action are all expressed in even powers of the QCD coupling. That is because only hard modes are 
present in the integrals. Odd powers will appear as soon as we admit modes of order $gT$. 

 The parameters in the $A_0$ sector are needed up to two loop accuracy~\cite{kajeff} and we give the result for $N_c=3$ and $N_f$:
\begin{equation} 
m^2_E =g(\mu)^2(1+{N_f\over 6})T^2(1+\Delta g^2(\mu))\nonumber
~\mbox{and}~\\
\lambda_E ={3g(\mu)^4\over {8\pi^2}}T(1-{N_f\over 9})(1+\delta g^2(\mu)).
\label{eq;paraelec}
\end{equation}

For $\bar\lambda_E$ see reference~\cite{bronoff}.
The coefficients $\Delta$ and $\delta$ depend logarithmically on the scale $\mu$
and for their explicit form see refs.~\cite{kajeff}~\cite{bronoff}.

The gauge coupling $g_E$ starts to run and in the $\overline{MS}$ scheme one finds
~\cite{huang}:
\begin{equation} 
g_E^2=g^2(\mu)T\{1+{g^2(\mu)\over{(4\pi)^2}}({22N\over 3}\log{\mu\over{\mu_T}}-{4N_f\over 3}\log{4\mu\over{\mu_T}})\}.
\label{eq:running}
\end{equation}

The parameter $\mu_T=4\pi T\exp{(-\gamma-{1\over{22}})}=6.742..T$ follows from the one loop renormalization of the $F_{ij}^2$ term through the effects of scale $T$~\cite{huang}. Eulers constant $\gamma$ equals $0.577214..$.
If one subtracts at this scale the renormalization effects appear only to two loop order and the coupling is then function of $T/\Lambda_{\overline{MS}}$:
\begin{equation}
{g_E
^2N\over{T}}={24\pi^2\over{11\log({6.742..T\over{\Lambda_{\overline{MS}}}})}}.
\label{eq:minsub}
\end{equation}

Quenched QCD lattice simulations give us the critical temperature in 
terms of the QCD scale~\cite{heller}:

\begin{equation}
{T_c\over{\Lambda_{\overline{MS}}}}=1.15\pm 0.05, N=3~\mbox{and}~ 
{T_c\over{\Lambda_{\overline{MS}}}}=1.23\pm 0.11, N=2
\end{equation}

~From the three dimensionful quantities in this Lagrangian we can~form two dimensionless quantities:
\begin{equation}
x={\lambda\over{g_E^2}}~\mbox{and}~y={m_E^2\over{g_E^4}}
\label{eq:xy}
\end{equation}

We have made our promise true that the scale $\Lambda_{\overline{MS}}$ only goes in through the running of the coupling. 

The dimensionless couplings $x$ and $y$ contain both $g^2(\mu)$. Eliminating 
the latter gives a very simple relationship between the former:
\begin{equation}
xy|_{4D}={2\over{9\pi^2}}(1+{9\over 8}x+O(x^2))~\mbox{for}~ N=2
\label{eq:nistwo}
\end{equation}
and 

\begin{equation}
xy|_{4D}={3\over{8\pi^2}}(1+{3\over 2}x+O(x^2))~\mbox{for}~ N=3.
\label{eq:nisthree}
\end{equation}

 This is the trajectory in the x-y plane of the 4D physics, to order $O(x^2)$.
We put $N_f=0$.
Remarkable is that it does not depend on the subtraction scale $\mu$!
The subtraction scale survives of course in the variable x but not in the
relation between $x$ and $y$. If this 
trend continues in higher orders, the series in $x$ is probably well convergent.

Of course the physics of this effective action is specific to the quark-gluon plasma. First the coupling should be sufficiently small, and the presence of the mass term indicates that the electric flux is screened. And indeed $m_E$ is to lowest order identical to the Debye mass $m_D$ since both equal the one loop static self energy at zero momentum.    The difference comes in the corrections. Whereas the corrections to $m_E$ are $O(g^2)$ due to the absence of soft modes, those to $m_D$ are 
$O(g)$ due to presence of soft modes. The Debye mass $m_D$ is a physical quantity.  $m_E$ is a parameter in the electrostatic action.

The mixed sector is known to one loop~\cite{chapman} up to six external legs. We lumped it into the term $\delta{\cal{L}}_E$. The reason for doing so is a question of accuracy. Already the superrenormalizable terms retained in  eq.(\ref{eq:estat}) do insure that the  error we make in calculating some observable $O$ with our electrostatic action is $O(g^4)$. This error to be numerically small constitutes  one of several constraints on the value of the coupling. It warrants the calculation of the mass and four point coupling to two loop order above.

Let's see how this accuracy comes about.  The argument is dimensional and based on the invariances of the reduced theory. These are the discrete spatial symmetries, 3D rotational and gauge invariance.    Including two extra spatial potentials on any of the terms in ${\cal{L}}_E$ you get six independent terms~\cite{chapman} (~from $F^3,(DF)^2, A_0^2F^2, A_0FDF$). A typical term reads:
\begin{equation} 
\delta{\cal{L}}_E\sim {g^2\over{T^2}} (DF)^2.    
\end{equation}
The square of the coupling appears because of the interaction of the stationary modes with the heavy modes. The scale $T$ is there for dimensional reasons.
The question is what this vertex is going to contribute. Irrespective of the observable in which it appears,  we can say that the covariant derivative $D$
concerns momenta in the effective theory of $O(gT)$. That gives an $g^4$ factor in front of the $F^2$ factor already present in the original Lagrangian
 ${\cal{L}}_E$, and provides the order of the relative error. 

The estimate is generic. It can be higher order for specific observables.
It motivates the two loop accuracy for $m_E$ and $\lambda_E$ above.

 Another constraint is the following. The cut-off for our theory is $2\pi T$. The Debye mass ($\sim gT$), a typical scale of our electrostatic theory, should then be smaller than this cut-off. This means $g\le 2\pi$, or $\alpha_s \le \pi$. 
  
 In terms of temperature scales:  if ${T\over{\Lambda_{\overline{MS}}}}\sim 2$ then our coupling
through eq.(\ref{eq:minsub}) with $N=3$ equals $g\sim 1.7$, consistent with the cut-off limit.

\subsection{Integrating out the electric screening scales}

On the other hand there is the scale $g_E^2=g^2T$.  If this scale is much smaller than the Debye mass, we can integrate out the scale $gT$ by integrating out the Higgs field $A_0$ from $S_E$. This necessitates the introduction of yet another cut-off
$\Lambda_M$ separating the scales $gT$ from $g^2T$.
That will lead to a new action $S_M$ with only the magnetic fields present. This magnetostatic action density reads:
\begin{equation}
{\cal{L}}_M=TrF_{ij}^2+\delta{\cal{L}}_M
\label{eq:mstat} 
\end{equation}   
\noindent
with a magnetostatic gauge coupling $g^2_M$.

This coupling is related to the electrostatic coupling $g_E$ through the renormalization of the magnetic gluon field strength:
$$F_{ij}^2\rightarrow (1+g_E^2Z)F_{ij}^2$$
 To one loop order the $A_0$ field is the only
field contributing. To compute the $Z$ factor we have to compute a diagram like in fig.(\ref{fig:12loop}d) with the wavy external legs the background magnetic
potential with momentum of $O(g^2T)$. There is also a tadpole-like diagram
contributing.

 Simple power counting gives a linear infrared divergence for the transverse result $Z$. Since the infrared in $S_E$ is cut off by $m_E$,  we expect parametrically $Z\sim g_E^2/m_E\sim g$. So odd powers of g are to be expected. For $g_M$ we get for three colours:
\begin{equation}
g^2_M=g^2_3(1-{g_E^2\over {16\pi m_E}})
\label{eq:magnel}
\end{equation}
for all reasonable couplings $g\le 1$ a small effect.

Using the magnetostatic action at scales $g^2T$ or smaller will induce an error $O(g^3)$ with respect to the results one would have got with the electrostatic action. 
Like in the previous subsection a generic estimate tells us for a typical term from the correction term in eq.(\ref{eq:mstat}):
\begin{equation}
\delta{\cal{L}}_M\sim{ g_E^2\over {m_E}^3}(DF)^2
\label{eq:magnell}
\end{equation}
The coupling $g_E$ describes the interaction between electric and magnetic modes, and $m_E$ the scale of the integrated degree of freedom $A_0$. Now $D\sim g^2T$ and the relative error is $O(g^3)$.

 This magnetic  action has no dimensionless couplings like the electrostatic one. 
At scales $g_M^2$ it's obvious we cannot form a small dimensionless number with the coupling $g^2_M$. So
the coupling in this theory is strong. A formal perturbation expansion of say the free energy gives from four loop order ($O(g_M^6)$) on powerlike infrared divergencies as naive power counting shows. Regulate with a mass $m$. Now, any free energy diagram with L
loops has a power $(g^2_M)^{(L-1)}$ in front of the integral. So the integral 
must give a result $m^{(4-L)}$ to get the correct dimension for the the free energy. For $L=4$ one  expects a logarithm in the ratio cut-off over mass, calculated recently~\cite{log}. For $L\ge 4$ we have linear or higher divergencies. For a superrenormalizable theory all logs containing the cut-off are contained in $L=4$.

 When one regulates these divergencies with a mass of $O(g^2_M)$ higher loop diagrams are all of order $g_M^6$ modulo logarithms. This is Linde's argument~\cite{linde}.

What this means is that the coefficient of the sixth order free energy is not perturbatively  calculable. We need non-perturbative input like the lattice. 

If one needs higher order effects then the term $\delta{\cal{L}}_M$
has to be expanded in the magnetostatic partitionfunction $Z_M$:
\begin{eqnarray}
Z_M & = & \int D\vec A\exp{\big(-S_M(A)-\int d\vec x \delta{\cal{L}}_M\big)}\\
    & = & \int D\vec A\exp{-S_M(A)} (1-\int d\vec x\delta{\cal{L}}_M+\ldots{....})
\end{eqnarray}

This gives an expansion for $-{1\over V}\log Z_M=ag^6T^3(1+bg+....)$
where a, b and higher order coefficients have to be computed on the lattice or any other non-perturbative method.

\section{Dimensional reduction at work}\label{sec:work}

We will treat a few examples of relevant observables in order of mounting complication.

 We will start with the spatial Wilson loop. Then the magnetic screening length $m_M$. Then its electric analogue, the  Debye mass and finally the spatial 't Hooft loop and the pressure. We want to calculate the first terms in the series up to and including the  term where the magnetostatic action enters for the first time. We just saw that this coefficient has to be computed from 3D lattice simulations. It turns out to dominate for any reasonable temperature, say from a few times $T_c$ till $10^5 T_c$. Then one compares to 4D lattice data to determine the remnant of the series. This remnant turns out to be small, typically on the order of 30$\%$ up to $T\sim 2T_c$ at least for Wilson loop and Debye mass. For pressure, magnetic screening length  and 't Hooft loop, this program is being pursued. 

\subsection{Spatial Wilson loop and magnetic screening mass: a window on the magnetic sector}

The spatial Wilson loop is given in terms of a spatial loop $L$, and a representation $r$ of $SU(N)$. The vector potential in this representation, $\vec A_r$,
 appears in the loop as:
\begin{equation}
W_r(L)=Tr{\cal{P}}\exp{i\oint_L g\vec A_r.d\vec l}.
\label{wilsonloop}
\end{equation}

As for the Wilson line, eq.(\ref{eq:wilsonline}), the exponential is path ordered and hence invariant against regular gauge transformations . This spatial loop should measure the magnetic flux of any fixed gauge field configuration, as suggested by its abelian analogue and Stokes theorem.
There is a useful version of Stokes theorem for the non-Abelian case~\cite{petrov}~\cite{kovner}. 

The thermal average of the spatial Wilson loop shows area behaviour with a surface tension $\sigma_r(T)$.
 \begin{equation}
\langle W_r(L)\rangle_T=\exp{-\sigma_r(T)A(L)+\ldots{...}}.
\label{wilsonloopaverage}
\end{equation}

The dots indicate perimeter terms. As far as the tension is concerned, it is very plausible~\footnote{ See the notes of Prof. Teper in this volume.} that it {\it only} depends on the number of quark minus the number of anti-quark representations constituting the representation $r$. This number is called the N-allity $k$ of the loop. Also the tension is {\it periodic} in $k$ modulo $N$. 
For the tension of the 't Hooft loop these properties are verified easily from its definition.

% This behaviour is made plausible by  the behaviour of the loop at zero temperature.
%At $T=0$ timelike and spacelike  loops are identical. In absence of dynamical quarks the timelike Wilson loop can have area behaviour because the timelike Wilson loop can be viewed as the correlation between a heavy charge $Q$ and its charge conjugate $\bar Q$. 

%In particular for the fundamental loop
%one has an area law. The string tension is due to the tension (energy per unit length) of a fluxtube emanating from the fundamental charge $Q$ and terminating on the charge conjugate charge $\bar Q$. Add to the fundamental charge a pair $\bar Q Q$, and to the charge conjugate as well. This means we are adding a fundamental loop and its charge conjugate. Clearly the area law will not change, the perimeter law will change, because the extra $Q\bar Q$ pair adds a pair of flux -anti-flux that annihilates. If we add $N$ charges $Q$  the N fluxes will break and form a baryon on one side and an antibaryon on the other side of the loop. So the dependence on N-allity $k$ modulo $N$ follows.

A useful corollary: a loop with N-allity $k$ ( a k-loop) and a $-k$-loop have by charge conjugation the same tension. So because of the periodicity also the $N-k$ loop has the same tension. 

%The spatial Wilson loop at high $T$ is obtained by continuously varying $T$ so these properties will not get lost.

For $N=2,3$ only one  tension results, because of charge conjugation and periodicity.

%%%%%%%%%%%%%%%%%%%%%%%%%%%figure%%%%%%%%%%%%%%%%%%%%%%%%%%%%%%%%%%%%%%%%%
\begin{figure}[htb]
\begin{center}
 \epsfig{bbllx=55,bblly=55,bburx=540,bbury=740,
         file=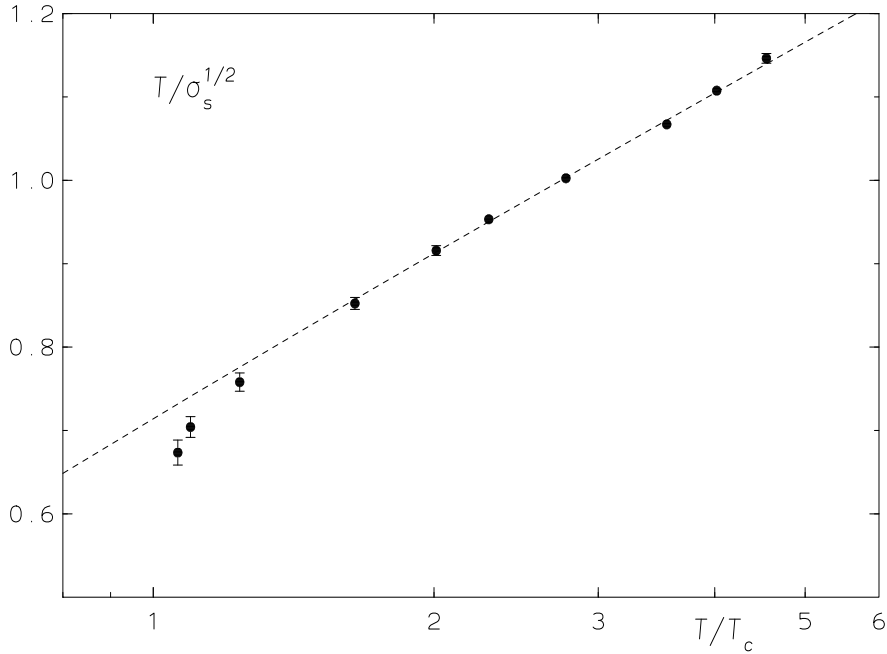,width=90mm,height=120mm,angle=-90}
\end{center} 
\caption{The temperature over the square root of the spatial string tension
 versus $T/T_c$ for $SU(3)$. The dashed line shows a fit 
according to a two loop scaling formula for the coupling, see text below eq.(\ref{eq:wtensionmagnetic}). From ref.\cite{karschtension}. }
\label{fig:string}
\end{figure}
%%%%%%%%%%%%%%%%%%%%%%%%%%%%%%%%%%%%%%%%%%%%%%%%%%%%%%%%%%%%%%%%%%%%%%%%
We now establish that the loop is a perfect window on the magnetic sector.
Its thermal average in path integral language is:
\begin{equation}   
\langle W_k(L)\rangle=\int DA_0 D\vec A W_k(L)\exp{-S(A)}/\int DA_0D\vec A\exp{-S(A)}.
\end{equation}

Integrate out all hard modes. They will  not contribute to the tension $\sigma_k$ of the loop, because the tension is due to correlations of the potential $\vec A$ over macroscopic distances.  That will leave us with $S$ replaced by $S_E$ on the r.h.s. of the average.
Since the  spatial loop contains only spatial potentials $\vec A$ we can
 integrate over $A_0$ to obtain $S_M$ from $S_E$. We arrive  for the tension at the average:
\begin{equation} 
\exp{-\sigma_k(T)A(L)}=\int  D\vec A W_k(L)\exp{-S_M(A)}/\int D\vec A)\exp{-S_M(A)}.
\label{eq:wloopmagnetic}
\end{equation}
The hard and electrostatic free energies $f_h ~\mbox{and}~ f_E$ drop out in the ratio.

The only dimensionful scale in the magnetostatic action is $g_M^2$. So
the tension, having dimension $\mbox{(mass)}^2$, can be written as:
\begin{equation}
\sigma_k(T)=c_kg_M^4(1+O(g^3)).
\label{eq:wtensionmagnetic}
\end{equation}

So the dominant contribution to the tension is entirely from the magnetostatic sector. 
In  figure (~\ref{fig:string}) you see a fit of the tension data to this parametric expression for SU(3).
The authors took for the magnetic coupling $g_M^2=g_E^2$, so neglected renormalization effects of the scale $gT$, which are a few percent at $T=2T_c$, see eq.(\ref{eq:magnel}). On the other hand they included two loop renormalization effects. Dropping those effects, and taking into account the uncertainty in the relation between $\Lambda_{\overline{MS}}$ and $T_c$ there is still consistency between data and the one loop formula eq.(\ref{eq:minsub}).

Notably the value of the tension at the critical temperature
is within errors equal to the tension at zero temperature. So the tension of the spatial Wilson loop
 does not change within errors in the hadron phase.

The conclusion is quite clear: down to temperatures a few times $T_c$, the loop
 behaviour is determined by leading order magnetic sector effects! These effects are embodied in the dimensionless number $c_{k=1}$. The rest of the T-dependence is through the hard-mode-running of the coupling, eq.(\ref{eq:minsub}).
The number $c_{k=1}$ is within errors equal to the purely 3D simulation
of the loop. 

The spatial Wilson loop measures in a sense to be specified later the magnetic flux in the system. The tension is flat from $T=0$ to $T=T_c$, according to the data.  In all of the confined phase the magnetic activity does not
change. 

Above $T_c$ it starts to grow like $g_M^4 T^2$. Apparently  beyond the transition the activity goes up, and comes, as the data tell us, entirely from the magnetostatic sector.

This window on the magnetic sector spurs an obvious question~\cite{giovannakorthals}: what is the dependence of the coefficient on $k$ and $N$? The lattice data by Teper and Lucini~\cite{luciniteper} are consistent within a percent to the simplest possible
picture for the 3D magnetic sector: that of a gas of almost free quasi-particles, in some sense ``static transverse gluons''. We'll come back to this in the last section.

\subsubsection{The magnetic screening mass}\label{subsubsec:magscreen}

The magnetic screening mass $m_M$ was introduced in section(\ref{sec:forces}) as the magnetic analogue of the Debye mass. It  gets its leading order contribution from
the magnetostatic sector, like the spatial Wilson loop.

\begin{figure}[h]
\begin{center}
\epsfig{figure=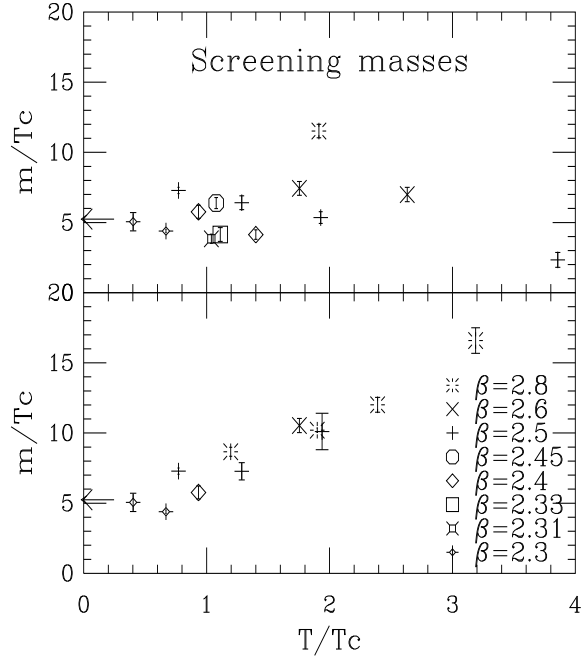,height=8.0cm,width=7.0cm}
\end{center}
\caption{Screening mass as a function of temperature, both in units of $T_c$,
as extracted from spatial (top) or temporal (bottom) 't~Hooft loops.
Below $T_c$ both coincide.
The arrow gives the mass of the scalar glueball at $T=0$. From ref.~\cite{defor}.}
\label{fig:screening_mass}
\vspace{-0.1cm}
\end{figure}

Four dimensional data
for SU(2) have been taken~\cite{rebbi}~\cite{defor}. But the numerics is much more involved than that for the Wilson loop. Qualitatively the data for the screening mass are compatible with the 
behaviour of the spatial Wilson loop tension. In the cold phase  its value is about twice  the lowest glueball mass. Beyond $T_c$ it starts to rise, as you can see from the  lower part in  fig.(\ref{fig:screening_mass}):
\begin{equation}
m_M=rg^2_M=rg^2(T)T
\label{eq:magmass}
\end{equation}
\noindent
 as one would expect from a mass in the 3d theory.

 Because the mass is high, the signal to noise ratio becomes small and numerical extraction becomes tedious.  

The 4D data are being improved~\cite{rebbi} for the region around $2T_c$, where we want to confront them with the 3d data.

Quantum numbers of the  magnetic screening  mass in SU(2) are given by 
$J^P=0^+$~\footnote{ Remember in SU(2) gauge theory all of the gauge-invariant sector has  positive charge conjugation. This is due to the pseudo reality of SU(2):
any element $u$ is equivalent through $\sigma_2 u\sigma_2=u^{*}$ to its complex conjugate. And the Pauli matrix $\sigma_2$ is precisely charge conjugation:
$A_{\mu}^{cc}=\sigma_2A_{\mu}\sigma_2$. As the gauge invariant sector involves integrating over charge conjugation, so is charge conjugation invariant.  We drop the label for $C$  altogether.}. 
%%%%%%%%%%%%%%%%%%%%%%%%%%%%%figure%%%%%%%%%%%%%%%%%%%%%%%%%%%%%%%%%%%%%%
%%%%%%%%%%%%%%%%%%%%%%%%%%%%%%%%%%%%%%%%%%%%%%%%%%%%%%%%%%%%%%%%%%%%%%%%%%%%
Precise  3D data for SU(2) are available from ref.~\cite{hartowe} and~~\cite{teper3d}. The 4D data are certainly not compatible with the lowest $0^+$ state. There are two recurrences, the highest of which (ref.~\cite{teper3d}, table 28) is compatible with the 4D data at $T=2T_c$ in fig.(\ref{fig:screening_mass}), up to 30\%. The situation is currently under investigation~\cite{hoelbling}.

\subsection{The Debye mass} 

%The spatial 't Hooft loop~\cite{thooft} is defined by a loop $C$ carrying a centergroup flux $\exp{ik{2\pi\over N}}$. This means the loop is given by any gauge transformation $V_k(C)$ with a discontinuity $\exp{ik{2\pi\over N}}$ across the minimal surface spanned by $C$.   

%Ex. Prove  the commutation relation of 't Hooft.

%Like for the Wilsonloop we can integrate out the hard modes. But we can integrate out the n=0 mode as long as the profile is itself
%$O(T)$. To one loop order we can do this with impunity. 
%%%%%%%%%%%%%%%%%%%%%%%figure%%%%%%%%%%%%%%%%%%%%%%%%%%%%%%%%%%%%%%%%%%%%%
\begin{figure}
\begin{center}
\includegraphics{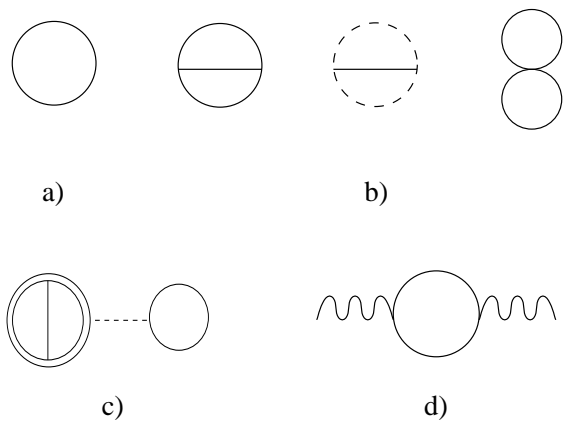}
\caption{(a) is the one loop contribution, and stands for gluon and ghost loop.(b) is the two loop contribution.(c) is the renormalization of the thermal Wilson line(double circle) inserted  (dotted line) into the one loop.(d) is the kinetic term with the loop being a gluon or ghost loop}
\label{fig:12loop}
\end{center}
\end{figure}

%%%%%%%%%%%%%%%%%%%%%%%%%%%%%%%%%%%%%%%%%%%%%%%%%%%%%%%%%%%%%%%%%%%%%%%%
%%%%%%%%%%%%%%%%%%%%%%%%%%%%%figure%%%%%%%%%%%%%%%%%%%%%%%%%%%%%%%%%%%%%%%
\begin{figure}[t]
\begin{center}
\epsfig{figure=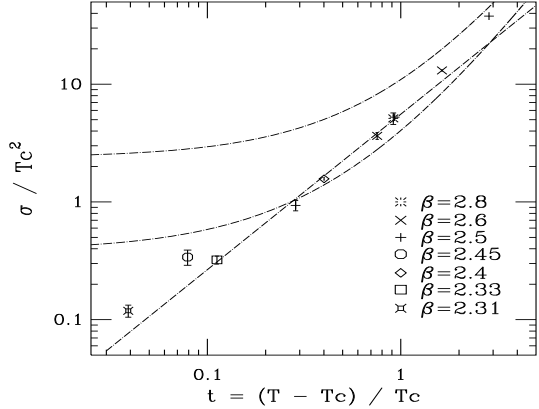,height=5.0cm,width=6.5cm}
\end{center}
\caption{'t Hooft loop tension $\rho_1$ for SU(2), in units of $T_c^2$, as a function of 
the reduced temperature $t$. The straight line is a power law fit to $t < 1$.
The fitted exponent is $1.32(6)$, to be compared with $2 \nu \approx 1.26$
for the $3D$ Ising model. The curves show the perturbative result,
to leading (upper) and next (lower) order, eq. (\ref{eq:result}) in the text. From ref.\cite{defor}.} 
\label{fig:sigma}
\vspace{-0.5cm}
\end{figure}
%%%%%%%%%%%%%%%%%%%%%%%%%%%%%%%%%%%%%%%%%%%%%%%%%%%%%%%%%%%%%%%%%%%%%%%%%

The Debye mass to next to leading order is given by:
\begin{equation}
m_D=({N\over 3}+{N_f\over 6})^{{1\over 2}}gT+{g^2N\over{4\pi}}T\big(\log({{N\over 3}+{N_f\over 6})^{{1\over 2}}\over g}+7.0\big)+O(g^3)
\label{eq:debyemasshighorder}
\end{equation}

The first term is the  result from the one loop self-energy discussed in section (\ref{sec:forces}).
Remember only hard modes contributed to one loop order.

The second term is only $O(g)$ smaller. The log term in the coefficient is
due to Rebhan~\cite{rebhan} and comes from scales between electric and magnetic ones.
The large number comes from all scales equal or smaller than the magnetic one,
$g^2_M$. It is non-perturbative and calculated by numerical simulation on the 
lattice~\cite{kajdebye}~\cite{hartowe}~\cite{hartlaine} for N=2 and 3. How is explained below.

A natural choice in 3d is the reduced version of the Wilson line: $P=Tr\exp{iA_0}$.
 Its imaginary part is the negative R parity channel. 
\begin{itemize}
\item
Exercise: Use the electrostatic action ${\cal{L}}_E$ in the correlator. \\
Show $<Im P(0)Im P(z)>_E$ has as dominant decay mode $\exp{-3m_Dz}$
\end{itemize}

The mass can be extracted from this correlator. In terms of the electrostatic parameters eq.(\ref{eq:xy}) it has a simple form:
\begin{equation}
m_D=m_E(1+d\sqrt{{1\over y}}+ O(x^2))
\label{eq:debyemassnonpert}
\end{equation}

We have to evaluate the correlator along the physics line eq.(\ref{eq:nisthree}).
To see that we can expect this form for the correction take the self energy correction to one of the three $A_0$ propagators contributing to $<Im P(0)Im P(z)>_E$.  Let the momentum of this propagator be $\vec p$. Schematically, the self energy insertion reads:
\begin{equation}
g_E^2\int{d\vec l\over {(2\pi)^3}}{(2l+p)^2\over{(l^2+m_E^2)((l+p)^2}}
\end{equation}

We are interested in the contribution of this integral to the Debye mass in the exponent $\exp{-m_Dz}$. So we have to evaluate the integral on-mass shell:
$\vec p^2+m_D^2=0$. That contribution is for dimensional reasons $\sim {1\over{m_E}}$. That,  together with the coupling constant factor $g_E^2$ gives indeed  ${g_E^2\over{m_E}}=\sqrt{{1\over y}}$. 

Here the value of the coefficient $d$ for N=3 is given.  Its  presence renders the next to leading term dominant till $T\sim 10^6 T_c$!

But what is even more so is that the remnant of the series in eq.(\ref{eq:debyemasshighorder}) converges well up to $T=2T_c$.
This follows from comparing to the full 4D lattice data~\cite{karschdebye}. 

And so the pattern here is the same as for the spatial Wilson loop: once the magnetic sector has contributed  to the series the rest is well convergent.

\subsection{Spatial 't Hooft loop}\label{subsec:thooftloop}

In this subsection we finally make good our promise to compute the first three orders in the tension of the 
spatial 't Hooft loop at high temperature. We met this tension in  section (\ref{sec:canoniczn}), where it came up as a ratio of twisted partition functions,
eq.(\ref{eq:initialrho}).

First we shall give a definition that makes clear its connection with the thermal Wilson line and with thermal Z(N) symmetry.

Consider a closed loop $L$ in the (x,z) plane, and define the spatial 't Hooft loop  $V_k(L)$ as a gauge transformation that has a discontinuity $z_k$ on the minimal surface spanned by the loop.

This definition tells us that the loop is a closed magnetic flux loop with strength $z_k$, as we defined in section (\ref{subsec:prolegozn}). 

 A simple realization of such a transformation is the solid angle $\omega_L(\vec x)$ with which $L$ is seen from a point $\vec x$. It jumps over $4\pi$ when  crossing the surface. So the corresponding   operator in Hilbert space is formed by taking Gauss's operator:
\begin{equation}
\tilde V_k(L)=\exp{i\int d\vec x{1\over g} Tr\vec E(\vec x).\vec D\omega_L(\vec x){Y_k\over {2N}}}.
\label{eq:thooftloopone}
\end{equation}
 Remember from section (\ref{sec:canoniczn}) that $Y_k=\mbox{diag}~(k,k,...,k,k-N,...k-N)$ with N-k entries k and k entries k-N, so that it generates the center group
element:
\begin{equation}
\exp{(i{2\pi\over N} Y_k)}=\exp{ik{2\pi\over N}}=z_k.
\end{equation}

What does a thermal Wilson line feel when it passes through the minimal area of the loop? 

The answer is simple: since the Wilson line represents a heavy test quark it
will pick up the Z(N) phase $z_k$.

 So the operator acts like a twist.  To recover the twisted partition functions $\Omega_{\vec k}$ of section (\ref{sec:canoniczn}) (with $\vec k=(0,k,0)$ for the case at hand), just extend $S(L)$ over the full $x,z$ cross section~\footnote{For the vortex correlation in the 2+1 dimensional theory in section (\ref{sec:canoniczn}) precisely the same holds true. Mutatis mutandis the reasoning of this section applies to that correlation as well.}. 
As promised in that section  we now compute the area law for the 't Hooft loop, when $T\ge T_c$:
\begin{equation}
<\tilde V_k(L)>=\exp{-\rho_k(T)S(L)}
\end{equation}
\noindent
and it is the tension $\rho_k(T)$ we are after.

\subsubsection{The strategy for computing $\rho_k(T)$}

Imagine the loop immersed in the plasma. Far away from the loop the value of the Wilson line is some fixed Z(N) value, as we learned in section (\ref{subsec:wilsonzndeconf}). We will take the value $P(A_0)=1$, but any other would have doe equally well. 

\begin{figure}
\begin{center}
\includegraphics[angle=270]{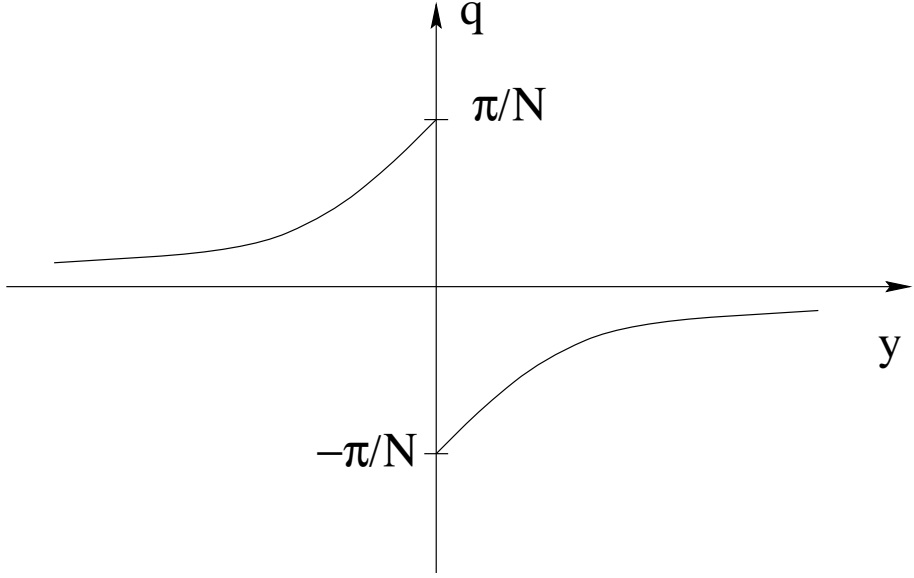}
\end{center}
\label{fig:wilsonjump}
\caption{The  Wilson line profile, that dominates the steepest descent calculation of the effective potential. The parameter $q$ is explained in the text.}
\end{figure}

The geometry of the problem is such that all profiles  orthogonal to the plane of the loop will be identical. Only near the border this is no longer true. We 
are only interested in the surface effects, not in the border effects.
As we said before the Wilson line will jump at the surface of the loop.
Such a typical profile is shown in fig. (\ref{fig:wilsonjump}).

The jump should be such, that the derivatives on both sides are equal. Once we have fixed these boundary conditions, all we have to do is to determine what profile minimizes the effective potential. This is what we are going to do in the next section.

\subsubsection{Effective potential for constant profile}

To  be specific  take SU(2) gauge theory. The key to computing the surface tension is the distribution function $E(\bar P)$ of the thermal Wilson line that we introduced in section(\ref{subsec:wilsonzndeconf}), eq.(\ref{eq:effaction}).  
We are interested in the logarithm  of this distribution function:
\begin{equation}
E(\widetilde P)=\int DA \delta(Tr\exp{iC}-Tr\overline{{\cal{P}}(A_0)})\exp{-S(A)}\equiv\exp{-L^3(V(q)+f)}
\label{eq:constraint}
\end{equation}

We have introduced the diagonal traceless matrix $C=\pi\mbox{diag}(q,-q)$ to parametrize $\widetilde P$. The free energy density $f$ normalizes the distribution. $L^3$ is the spatial volume.  
The distribution function has two peaks, one at P=1, one at P=-1. 
So the effective action $V(q)$ has  minima at $q=\mbox{integer}$. Even (odd) values correspond to $P=1 (-1)$.

 The constraint in eq.(\ref{eq:constraint}) tells us that the potential $A_0$ fluctuates around the background matrix $CT$. So this matrix will be present in the Feynman rules as a background field. This is familiar.
What is less familiar is that the background gets renormalized through  the renormalization of the thermal Wilson line. This is an effect coming in from two loop order on, and is shown in fig. (\ref{fig:12loop}c). Only the sum of the two loop diagrams b) and c) is gauge independent.

The covariant derivative is $D_0(A_0=TC)$.   Such a derivative is diagonal in the Cartan basis for the fields $A={CT\over g}+Q^+\tau^-+Q^-\tau^++Q^3\tau^3$. The Pauli matrices $\vec \tau$ relate to $\sqrt{2}\tau^{\pm}$ as $\tau_1\pm i\tau_2$.

So the Feynman rules consist of the 
Euclidean ones, except that $p_0$ is replaced by $p_0\pm 2\pi Tq$ when it is the Matsubara frequency of $Q^{\pm}$. Hence it appears in propagators of $Q^{\pm}$ and in vertices when it pertains to an outgoing $Q^{\pm}$ line.

 The  one loop computation (see fig.(\ref{fig:12loop}a) is the calculation of the determinant of the quadratic part in the quantum fields $Q$ of the action. In Feynman gauge you get:
\begin{equation}
V(q)+f={1\over{L^3}}\log\det(-D(C)^2).
\label{eq:logdet}
\end{equation}
The choice of gauge does not matter as is shown explicitely in Pisarski's lectures. $D(C)^2$ is the covariant Laplacian.

There is a well-known identity that relates $\log\det M=Tr\log M$ for any matrix $M$.
Then, in the  momentum basis you obtain in terms of the Fourier transformed Laplacian for
 $V(q)+f$:
\begin{equation}
T\sum_n\int {d\vec l\over{(2\pi)^3}}\big(\log((2\pi T (n-q))^2+\vec l^2)+
\log((2\pi T (n+q))^2+\vec l^2)+\log((2\pi T n)^2+\vec l^2)
\label{eq:su2pot}
\end{equation}
with the contributions from $Q^+, Q^-, Q^3$ integrations written out explicitely.
If we sum over all values of n, clearly the result will be periodic in $q$ mod 1 and even in q. This is of course a consequence of the thermal Z(N) symmetry. In fact one gets:
\begin{equation}
V(q)={4\over 3}\pi^2T^4q^2(1-|q|)^2
\label{eq:potential}
\end{equation}

As long as $qT=O(T)$ the integration over $\vec l$ is hard. For $q=O(g)~\mbox{or}~q=1-O(g)$ we have for $n=0~\mbox{or}~n=-1$ soft momentum contributions.
\subsubsection{Varying profile and gradient expansion}
Till now we computed the potential as if the profile $q$ were constant. But
what we are really after is the profile of $q$ as a function of $y$ in between the values $P=1$ and $-1$ at $y=\pm\infty$. 
That means we have to include the kinetic term $K=({T\over{g^2}}\partial_yC)^2$ to our potential $V$ to get the tension:
\begin{equation}
U(q)= \int_{-\infty}^{\infty} dy(K+V)
\end{equation}  

So we tunnel through the potential mountain $V(q)$ from $P=1$ to $P=-1$.

We swept a little problem under the rug. We forgot the
contribution from the one loop potential due to the gradient in $q(y)$! This however is of higher order as we will see shortly.

 To find the tension you have to minimize $U$
with the boundary condition that the wall is between P=1 and P=-1 regions.
This is done by the method of completing the square: 
\begin{equation}
U(q)=\int dy\big( (K^{1/2}-V^{1/2})^2+2(KV)^{1/2}
\label{eq:actionmin}
\end{equation}

The integration of the second term over y can be replaced by an integration over q, using the chain rule for $\int_{-\infty}^{\infty}dy K^{1/2}\sim\int_{-\infty}^{\infty}dy\partial_y q=\int_0^1dq$:
\begin{equation}
\int dy 2(KV)^{1/2}={2\over g} (8\pi^2T^2)^{1/2}\int_0^1 dq V^{1/2}
\label{eq:tension}
\end {equation}
which gives a number {\it{independent}} of the profile $q$!

Hence $U(q)$  in eq.(\ref{eq:actionmin}) is minimized for that profile q that obeys the equations of motion 
\begin{equation}
K^{1/2}-V^{1/2}=0.
\label{eq:eqnofmotion}
\end{equation}

The tension of the loop is then given by eq.(\ref{eq:tension}) and using the 
result (\ref{eq:potential}) for the potential one gets:
\begin{equation}
\rho_1(T)={4\pi^2\over{3\sqrt{6g^2}}}T^2(1-c_2~\alpha_s(T)+...)
\label{eq:result}
\end{equation}
Two loop corrections have been computed~\cite{bhatta} from the graphs in fig.(\ref{fig:12loop}b, c ~and~d), with the result $c_2=2.0682...$. 

What is the typical width of the Wilson line profile? Just look at the 
equations of motion eq.(\ref{eq:eqnofmotion}). Write them out with eq.(\ref{eq:potential}):
\begin{equation}
\partial_y q=m_Dq(1-|q|)
\label{eq:motion}
\end{equation}

\noindent 
with $m^2_D={2\over 3}g^2T^2$, the lowest order Debye mass from $N=2$. So it is the Debye mass that governs the width of the wall.  
This also answers the question about gradient terms. They are there, but are $O(g^2)$ compared to $V(q)$.

 Only hard momenta do contribute to this order, as you can see by computing
the contribution to $\rho_1$ from the wings of the profile near $P=0,1$.

The lattice simulated  tension~\cite{defor} in SU(2) is shown in fig.(\ref{fig:sigma}). For $T\sim 2T_c$ lattice data and perturbative prediction do agree reasonably well. 

The cubic corrections are now known~\cite{giovannakorthals} and  add a positive contribution. They are soft modes on the scale of the Debye mass. As we will see the same pattern shows  in the pressure, calculated with the same graphs as in fig.(\ref{fig:12loop}),
but with the background $q$ set to zero. The three loop contribution is in progress. 
 
Magnetic modes will contribute  through the next to leading order of the Debye mass (see eq. (\ref{eq:debyemasshighorder})) and have not yet been computed.

For general N and strength k of the loop one finds to one and two loop order 
a remarkably simple scaling law in k~\cite{giovannakorthals}:
\begin{equation}
\rho_k(T)={k(N-k)\over{(N-1)}}\rho_1(T)
\label{eq:kscaling}
\end{equation}
with $\rho_1(T)={4\pi^2\over{3\sqrt3g^2N}}(N-1)(1-1.0341..\alpha_s~N+O(g^3))$,
periodic in the strength  k mod N.

In one loop order this is a simple consequence of additivity of the potentials
for the various colour modes and of the tunneling path being along the
$Y_k$ direction. Let us look into that in more detail.

 One tunnels from 1 to $z_k$ by the path $qY_k$, from q=0 to q=1, so the diagonal background matrix is $C={2\pi q\over N}Y_k$. 
The operator whose determinant we have to compute is $-D(C)^2$~~\footnote{We drop
all indices exept colour indices.}.

Let us adopt the Cartan basis for the fluctuating potentials $Q$ in NxN matrix notation, like we did before for SU(2). This is a colour basis in which the effect of the profile is  diagonalized in the Laplacian $-D^2(C)$.

 Remember the profile $C$ is diagonal by construction, with  diagonal elements $C_i$, $i=1,....,N$. The profile appears only in the covariant derivative $D_0(C)$. There are 
diagonal fluctuations $Q^d$, that to one loop order do not contribute any $C$ dependence:
\begin{equation}
D_0(C)Q^d=\partial_0Q^d+iT[C,Q^d]=\partial Q^d.
\end{equation}

Then there are off-diagonal fluctuations $Q^{ij}$ ($1\le i\neq j\le N$) with
$Q^{ij}$ only non-zero  in the (ij) entry. For those you have:
\begin{equation}
D_0(C)Q^{ij}=\partial_0 Q^{ij}+iT[C,Q^{ij}]=\partial_0 Q^{ij}+i(C_{i}-C_{j})Q^{ij}
\label{eq:partialderiv}
\end{equation}
\noindent and we have diagonalized the Laplacian. The background field  $C$ comes in through the  diagonal elements  of the {\it{adjoint}} representation of $C={q\over N}Y_k$. They equal 0, or $\pm q$ up to a factor $2\pi$.

First we write the contribution to the $C$ dependent part in the potential
for a fixed combination (ij):
\begin{equation}
V_{ij}(C_i-C_j)=T\sum_n\int {d\vec l\over{(2\pi)^3}}\log( T (2\pi n+C_i-C_j))^2
\label{eq:fixedij}
\end{equation}

 The reader will recognize the form of the potential for SU(2),
  eq.(\ref{eq:su2pot}).

By changing the sign of the summation over Matsubara modes, it follows that $V_{ij}(C_i-C_j)$ is even in $C_i-C_j$. 

Remember that on the path $C={2\pi q\over N}Y_k$ the profile in its adjoint representation equals $|C_i-C_j|=2\pi q$ , or $0$. 
 So the $C$-dependent part of the full potential is obtained by multiplying $V_{ij}$ with the number of off-diagonal  modes  (ij) with
a  non-zero eigenvalue $C_i-C_j$ .

  This number is determined from $Y_k=\mbox{diag}(k,...,k,k-N,..k-N)$ by taking all differences of elements. There are $k$ elements with value $k-N$, and $N-k$ elements with value $k$. Hence there are $2k(N-k)$ ways of picking a non-zero  combination. We obtain for the full  potential:
\begin{equation}
V(q)=2k(N-k)(V_{ij}(2\pi q)-V_{ij}(0))=k(N-k){4\pi^2\over 3}T^4 q^2(1-|q|)^2.
\label{eq:fullpot}
\end{equation}

The second equality follows from eq.(\ref{eq:su2pot}) and eq.(\ref{eq:potential}).

For the kinetic term one has the identity $Tr(\partial_yC)^2={1\over {2N}}\sum_{i,j} {\partial_y(C_i-C_j)}^2$, valid because $Tr C=0$. So the same counting as above applies to the kinetic term:
\begin{equation}
K={T\over g}^2Tr(\partial_y C)^2=k(N-k){4\pi^2 T^2\over {g^2N}}(\partial_y q)^2.
\end{equation}

Then eq.(\ref{eq:kscaling})
 follows from the minimization of $K+V$.
\begin{itemize}
\item
The stability group of $Y_k$ is $SU(N-k)\times SU(k)\times U(1)$.  Determine the dimensionality of the coset space $SU(N)/SU(N-k)\times SU(k)\times U(1)$. Explain the masses appearing the diagonal and off-diagonal propagators $-D^{-2}(C))$.
\end{itemize}

Some comments on the 't Hooft loop tension, eq. (\ref{eq:kscaling})
\begin{itemize}
\item
The result for the tension is $O(N)$ in the large N limit, in contrast to the tension of the Wilson loop. The latter is $O(1)$ as follows from the well-known
index loop counting, valid to all orders in perturbation theory. 
\item
The absence of a sacred cow is noteworthy in eq.(\ref{eq:kscaling}). For k of order 1 the result has corrections $O(1/N)$! A priori we would have expected corrections
of $O(1/N^2)$ in a theory with only gluons, like  in the pressure.  This anomalous correction is a simple consequence of the counting:  not  $N^2-1$ gluons, but only $2k(N-k)$ contribute to the tension, whereas they all do contribute in equal amount to the pressure.
 
\end{itemize}

Physically eq.(\ref{eq:kscaling}) is understood as being due to the flux of the  screened quasi-particles and will be discussed in section (\ref{sec:quasihooft}).

\subsection{Pressure}

\begin{figure}[tb]

\centerline{~~\begin{minipage}[c]{7.5cm}
    \psfig{file=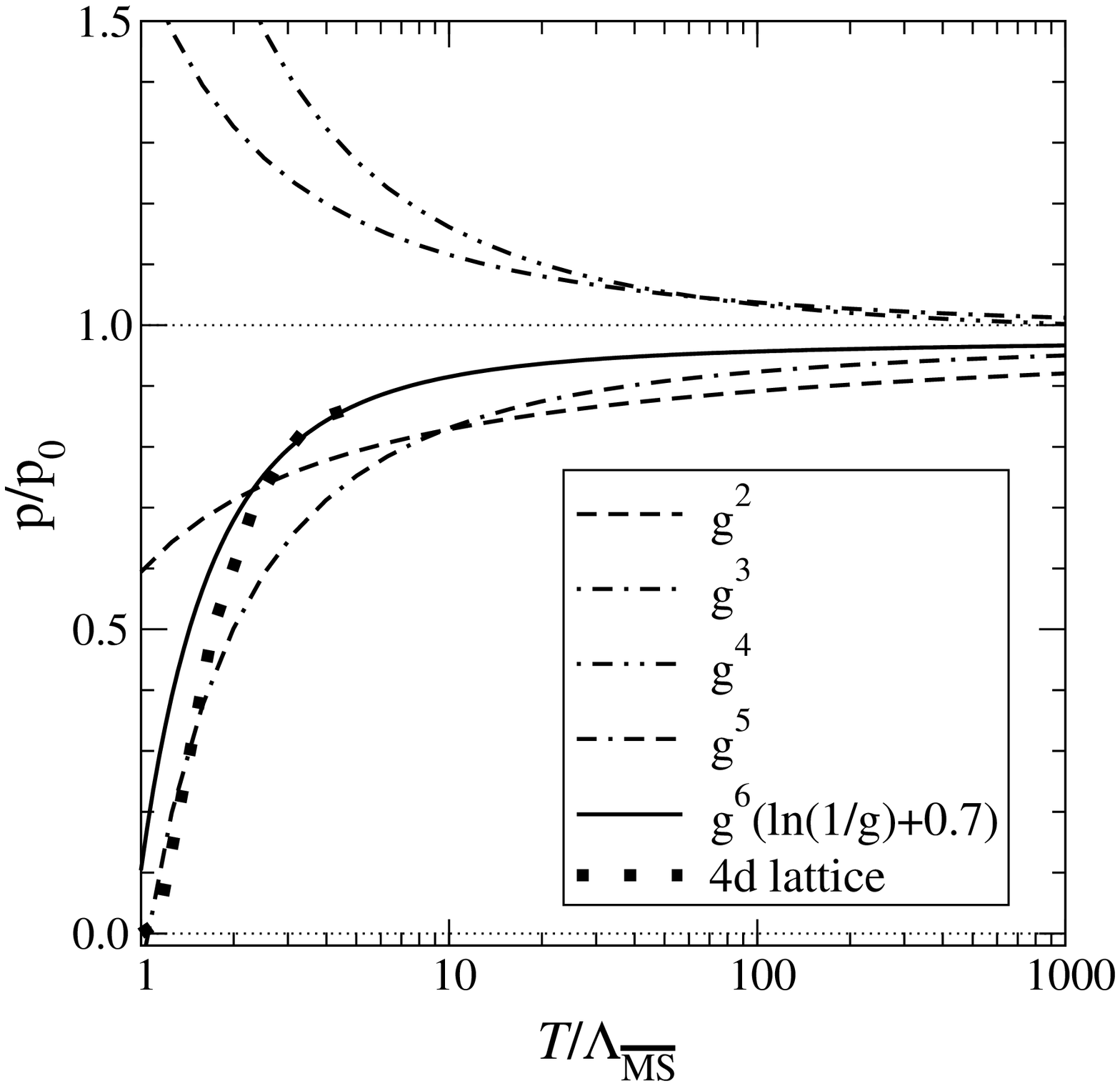,angle=0,width=7.5cm} \end{minipage}%
    ~~~~~\begin{minipage}[c]{7.5cm}
    \psfig{file=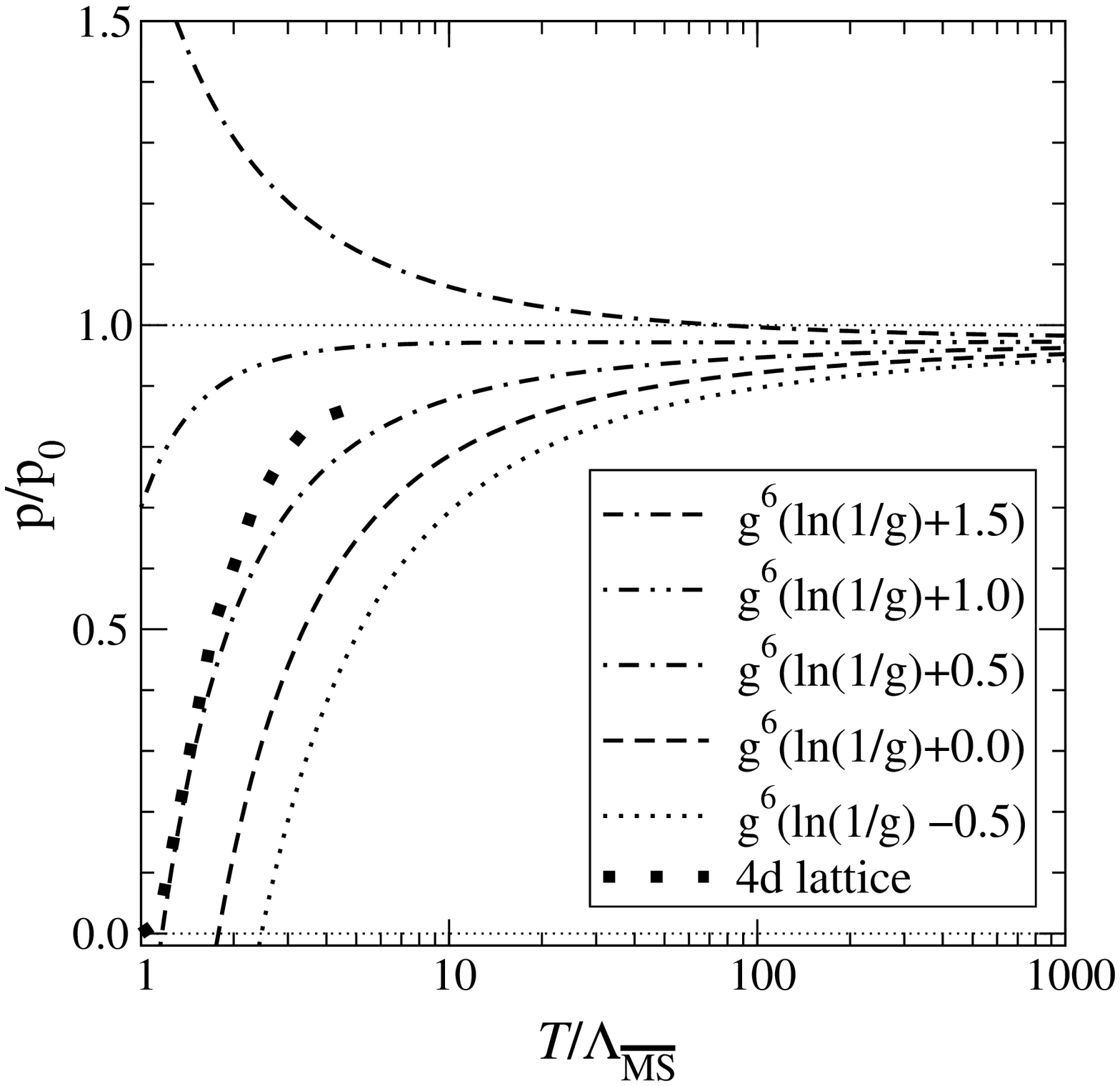,angle=0,width=7.5cm} \end{minipage}}

\vspace*{0.5cm}

\caption[a]{Left: perturbative results at various orders, including
${\cal O}(g^6)$ for an optimal constant. Right: the dependence of the 
${\cal O}(g^6)$ result on the (not yet computed) constant, which
contains both perturbative and non-perturbative contributions.
The 4d lattice results are from~\cite{boyd}. From ref.~\cite{log}.}

\label{fig:pert_g6}
\end{figure}

In fig.(\ref{fig:pert_g6})  you see the pressure
as measured by lattice simulation for three colours. It is compared to the  analytically computed~\cite{log} terms in perturbation theory up and including $g^6\log({1\over g})$. The series starts with the contribution from the gluons as free quasi-particles: the Stefan-Boltzmann gas with pressure $p_0={8\pi^2\over{45}}T^4$ as computed from fig.(\ref{fig:12loop}a) .  The interactions between the gluons give the two-loop contribution as in fig.(\ref{fig:12loop}b). The contribution of the Debye-screened
gluons is the dominant one from ${\cal{L}}_E$ in eq.(\ref{eq:estat}). It equals  the logarithm of the  determinant of the Higgs modes with mass $m_E$.  Using
dimensional regularization

%%%%%%%%%%%%%%%%%%%%%%%%%%%%%%%figure%%%%%%%%%%%%%%%%%%%%%%%%%%%%%%%%%%%%%%%%%%
%%%%%%%%%%%%%%%%%%%%%%%%%%%%%%%%%%%%%%%%%%%%%%%%%%%%%%%%%%%%%%%%%%%%%%%%%%%%

\begin{equation}
-{1\over 2}(N^2-1)\int {d\vec l\over{(2\pi)^3}}\log(\vec k^2+m_E^2)={\Gamma(-{3\over 2})\over{16\pi{{3\over 2}}}}m_E^3
\label{eq:cubic}
\end{equation}

This being positive and large with respect to the two loop contribution (like in the case of the 't Hooft loop) spoils the convergence of the series except for academically high temperature. The two next orders $g^4$ and $g^5$
undo in part the effect of the cubic term. But it is clear that perturbation theory has little predictive power in the region $2T_c~\mbox{to}~100 T_c$, as it varies there over more than 20\%!

 To put the calculation of the contributions of order higher than three in perspective and to see how the different scales come in, we recall once more
the hierarchy of scales, cut-offs $\Lambda$ and reduced actions needed to compute the pressure:
$$T>>\Lambda_E>>gT>>
   \Lambda_M>>g^2T$$.

The pressure is normalized by $p_0$ and consists of three parts:
$${p\over{p_0}}=p_h+p_E+p_M$$
The hard modes are cut-off in the infrared by $\Lambda_E$ and equal $p_h$.
Schematically we get:
$$p_h=1+g^2+g^4\log{T\over{ \Lambda_E}}+g^4+g^6log{T\over{\Lambda_E}}+g^6+..$$

All powers of the coupling are even. The short distance scales (larger than $T$)  are absorbed in the running coupling, eq.(\ref{eq:running}). The cut-off $\Lambda_E$ appears only in logarithms.
The electric mode contributions are computed with ${\cal{L}}_E$ and give $p_E$:

$$p_E=g^3+g^4\log{\Lambda_E\over{m_E}} +g^4+g^5+g^6\log{\Lambda_E\over{m_E}}+g^6\log{m_E\over{\Lambda_M}}+g^6+..$$
The dominant cubic term was computed in eq.(\ref{eq:cubic}). We can expect
 logarithms of the two ratios of the three scales $m_E$, $\lambda_E$ and  $\Lambda_M$ in the electrostatic action.

Finally the  magnetic contribution is computed with ${\cal{L}}_M$:
 $$p_M=g^6\log{\Lambda_M\over {g^2_M}}+g^6+...$$

We only put in the obvious dependence on the parameters in the electrostatic and magnetostatic actions. There are three comments:
\begin{itemize}
\item
All terms shown are perturbatively calculable, except the last one in $p_M$.
\item
All perturbatively calculable terms have been computed~\cite{zhai}, except for the $g^6$ terms. In particular the log's are known by now~\cite{log}.
\item
All dependence on the cut-offs cancels, as expected.
\end{itemize}

Clearly this is quite a calculational performance!

So at this point one expects that the miracle of the Debye mass may materialize:
compute the  non-perturbative term of $O(g^6)$ with ${\cal{L}}_M$ on the lattice, and the perturbative term of $O(g^6)$. Does the ensuing series stabilize down to $\sim 2T_c$?

 To see whether this may work at all,  values for the sum of the yet unknown non-perturbative and perturbative coefficients have been put in the right hand figure (\ref{fig:pert_g6}). Clearly there is a window where lattice data do connect smoothly to
the series.

\section{Flux of quasi-particles as seen by spatial Wilson and 't Hooft loops}\label{sec:quasihooft}

The idea of quasi-particles is to leading order in the pressure embodied by the Stefan-Boltzmann term. It counts all the degrees of freedom of the gluons, colour and spin.

What we want to argue below is that specific degrees of freedom of the gluons, namely their flux, are clearly seen in the 
behaviour of the spatial ~~'t Hooft loop, especially the scaling law in the strength of the loop. 
Then we will look for a similar effect in the Wilson loop, but now with quasi particles that carry magnetic flux.

\subsection{Gluon flux and the 't Hooft loop scaling law}

We found in perturbation theory including two loop order:
\begin{equation}
\rho_k(T)={k(N-k)\over {(N-1)}}\rho_1(T).
\label{eq:scaling}
\end{equation}

Below we will show that the factor $k(N-k)$ is the number of gluons in the adjoint multiplet  with a fixed value $N$ for the charge characterizing the strength $k$ of the loop.
Essential is  the physical meaning of the loop: it measures the colour electric flux in the plasma. 
The loop, eq.(\ref{eq:thooftloopone}), can be rewritten as a sheet of electric dipoles:
\begin{equation}
V_k(C)=\exp{{i4\pi\over N}\int d\vec S.Tr\vec EY_k}
\label{eq:fluxloop}
\end{equation}
\noindent
and measures the colour electric flux going through the loop.

Expression (\ref{eq:fluxloop}) is a dual Stokes law: remember the original definition
of the 't Hooft loop is that of a loop of a Dirac magnetic flux. But there is no 
magic about  (\ref{eq:fluxloop})!
 Both definitions can be shown to have the same commutation relation with the Wilson loop
\begin{equation}
V_k(C)W(C')V_k(C)^{\dagger}=z_k^{n(C,C')}W(C')
\label{eq:commutationrel}
\end{equation}
$n(C,C')$ being the number of times $C$ and $C'$ loop each other.

Hence 
the product $\tilde V_k(C) V^{\dagger}(C)$ does  commute with the Wilson loop. 

So the product is a regular gauge transformation  and acts
 as the unit operator in the physical Hilbert space. Both the representation eq.(\ref{eq:thooftloopone}) and the flux representation eq.(\ref{eq:fluxloop}) are  identical there.
 
\begin{itemize}
\item

Show that $V_k(C)$ has the 't Hooft commutation relation eq.(\ref{eq:commutationrel}) with a spatial Wilson loop $W(C')$.   (Hint: use the canonical commuation relations between field strength $\vec E$ and potential $\vec A$ in the Wilson loop.)
\end{itemize}

Consider a gluon close to the minimal area of the loop. As its flux is screened  it needs to be within the screening length $l_D$ to shine its flux through the loop~\footnote{This is a simplification. Also gluons farther out can contribute, giving a change in the  overall factor of our final result.}.  The gluon is in the adjoint representation so its charge $Y_k$  is either 0 or $\pm N$ as you can see from the differences of eigenvalues of $Y_k$. There are then obviously
 $2k(N-k)$ gluon species in the adjoint representation with this charge $\pm N$.
Each one individually shines  flux through the loop. This flux equals $\pm N/2$, the other half of the flux is lost on the loop. That means that such a gluon will 
give a contribution to the loop of 
\begin{equation}
V_k(C)|_{one ~gluon}=\exp({\big(i{2\pi\over N}(\pm {N\over 2})\big)}=-1.
\end{equation}

Assume that the distribution $P(l)$ of gluons in the slab of thickness $l_D$ containing the loop is
independent of the species and say Poissonian: $$P(l)={1\over{l!}}{\bar l}^{~l}\exp{-\bar l}.$$

Here  $\bar l$ is the average number of gluons of that species in the slab.
Then the average of the loop over just one species is:
\begin{equation}
<V_k(C)>|_{one~ species}=\sum_l(-)^{~l}{1\over{l!}}{\bar l}^{~l}\exp{-\bar l}=\exp{-2\bar l}.
\label{eq:poissonav}
\end{equation}

If the contribution of  all  $2k(N-k)$ species is independent the result becomes:
\begin{equation}
<V_k(C)>=\exp{-4k(N-k)\bar l}.
\end{equation}
Now $\bar l=2l_DS(C)n(T)$. n(T) is the density of a specific gluon species and an area law results with the tension:
\begin{equation}
\rho_k(T)=8k(N-k)l_Dn(T)
\label{eq:scalingrho}
\end{equation}

This formula represents the physical raison d'\^etre of the tension.  It is due to
the screening of the electric flux of the gluons, their density and their degeneracy with respect to the charge $Y_k$ characterizing the strength of the loop. 
The Poisson distribution function is not essential to this result.  Any thermodynamic distribution function, that  is peaked around the average $\bar l$ with a width $\overline{l^2}-\big(\bar l\big)^2$ proportional to $\bar l$, will give an area law. It is this proportionality constant that will appear in eq.(\ref{eq:scalingrho}). But it does not depend on the strength $k$ and will drop out in ratios.

 Hard gluons with momentum of order T,
will have a photon like 
 distribution function 
$$P(l)\sim(\bar l/(1+\bar l))^l.$$
 As the reader can easily check, it has a variance $\overline{l^2}-\big(\bar l\big)^2=\big(\bar l\big)^2+\bar l$, so fluctuations of order 1. Not surprisingly it   does  not give an area law as you can find out by plugging it in eq.(\ref{eq:poissonav}).  Small fluctuations are essential for the area law.

\subsection{Magnetic flux and k-scaling of the  spatial Wilson loop}\label{subsec:wilsonkscaling}

The question is now: what  about similar  ratios of  the spatial Wilson loop?

We convinced ourselves in section (\ref{sec:work}) that the  leading contribution to its tension came from the static magnetic sector. This sector is populated with static transverse gluons, i.e. gluons with a static magnetic field screened in 
some non-perturbative way with a screening mass $m_M$. The  screening mass 
is inversely proportional to the coupling $g_M^2$ on dimensional grounds.
More precisely, in our discussion of the magnetic screening in section (\ref{sec:work}) we noticed that the proportionality constant was large with respect to other scalar masses, because of its unnatural parity.

 In the context of our model
for the 't Hooft loop tension, it is interesting to see what the ratio of the tension of the Wilson loop
to magnetic screening mass is according to the lattice~\cite{teper3d}:
\begin{equation}
{m_M\over{\sqrt{\sigma}}}=8.15(15)~~~~\mbox{SU(2)}
\label{eq:small}
\end{equation}
 
Suppose  that the screening length $l_M=m_M^{-1}$ of our magnetic quasi-particles is much smaller than the mean distance between them: 
\begin{equation}
n_M l_M^3\ll 1. 
\end{equation}

Both are parametrically equal to $g_M^6$. There must be a dynamical reason for this ratio to be small.
It supposes the magnetic screening of the quasi-particles is so efficient that they constitute a gas of approximately free lumps.  

If the latter is true, then the reasoning in the previous section would, mutatis mutandis, give for the tension of the Wilon loop:
\begin{equation}
\sigma=c~l_Mn_M(T)
\label{eq:density}
\end{equation}

\noindent where c is some numerical constant depending on your preferred probability
distribution.
So this relation turns the numerical result for the screening length eq.(\ref{eq:small}) into a numerical result for the density of screened magnetic quasi-particles:
\begin{equation}
c~n_Ml_M^3=0.015....
\end{equation}
This means one quasi-particle in about seventy screening volumes if c=1,
an a posteriori justification for the model. For the Poison distribution
this density is even smaller. 

This is for SU(2). 

Remember that for SU(3) and higher groups the magnetic screening was related to $0^{--}$ states. From Teper's work~\cite{teper3d}   the lowest  $0^{--}$ mass in units of the string tension  for SU(N) ($N\ge 3$)) gives:
\begin{equation}
n_Ml_M^3=0.028....(1-(1.64...)/N^2+O(N^{-4}))
\end{equation}

Although both screening mass and  square root of string tension are parametrically the same, their ratio is for dynamical reasons small!
So we may say that the quasi-particles are to a good approximation free.
How good will become more clear at the end of this section. 

So having taken courage we now turn to a flux representation of the spatial Wilson loop~\cite{petrov}~\cite{kovner}. In fact, if we want to find out about the strength or N-allity k of
the loop, we would expect in analogy with the 't Hooft loop, eq.(\ref{eq:fluxloop}), a magnetic dipole distribution projected on the k-charge:
\begin{equation}
W_k(C)\sim \exp{i{g\over N}\int_{S(C)} d\vec S.Tr\vec BY_k} 
\label{eq:trialwilson}
\end{equation}
\noindent
with the magnetic field strength $\vec B$ replacing its electric counterpart
$\vec E$. 
 This is incorrect for at least one reason: as discussed in section (\ref{sec:work}) the Wilsonloop with N-allity k defines a periodic tension $\sigma_k$ only after averaging. Now the left hand side of the equation is periodic.
So the equation must be considered in thermal-averaged form. There is much more to say about this formula!  For a derivation of a slightly different form see ref.~\cite{petrov}.

But if we accept this flux representation on the basis of its analogy with the 't Hooft loop, then the same reasoning as for the 't Hooft loop  applies  and one concludes that 
the tension $\sigma_k$  scales as its electric analogue:
\begin{equation}
\sigma_k(T)={k(N-k)\over {(N-1)}}\sigma_1(T)
\end{equation}
%Note that in the context of statistical screening this assumption is  wrong! T%ake the screened gluons with density $n(T)\sim T^3$ and screening length $l_D\%sim {1\over {gT}}$.  Clearly $n(T)l_D^3\ge 1$! 
%The point is that magnetic screening is already present at zero temperature,
%and hence different from statistical screening.
%So we have an octet (in general adjoint) representation of these lumps. 
%The second part of the  analogy is the flux representation for the Wilson
%loop.  

All we have to do now is to hear the verdict of the lattice;
the ratios found by simulation~\cite{luciniteper} are close- within a percent
for the central value-:
$$SU(4):\sigma_2/\sigma_1=1.3548\pm 0.0064$$
$$SU(6):\sigma_2/\sigma_1=1.6160 \pm 0.0086;\sigma_3/\sigma_1=1.808\pm 0.025$$

The results are that precise, that you see a two standard deviation, except for the second ratio of SU(6). As we said, magnetic quasi-particles are dilute but only approximately free. 

There is a less precise determination of the ratio $\sigma_2/\sigma_1=1.52\pm0.15$ in SU(5)~\cite{meyer}. But the central value is within 1 to 2\% of the 
predicted value $3/2$.
In conclusion, lattice results are quite encouraging. More simulations are under way~\cite{hoelbling}.

\section{Epilogue}
We highlighted some of the salient problems in the critical region of
deconfined and chirally restored QCD. Universality works well, when it
is defined at all. 

In the region above $2T_c$ we studied the  indications that perturbation theory works well if the excitations of the magnetic sector are included.
These excitations, the magnetic quasi-particles (``magnons''),
may be considered as screened  quasi-particles. 3d simulations suggest  that their screening  is so  strong that they form approximately independent lumps. 
This idea is strikingly vindicated by the dynamically small ratio of magnetic screening over string tension.  

 The idea  works quite well in the Wilson loop sector. And the result for the Debye mass again  supports this idea; including   the magnetic quasi-particles corrects a result which was off by more than 
O(1) without!
The non-perturbative part of the pressure to order $g^6$ is the partial pressure of 
the magnetic gluon gas.

 Unfortunately our arguments are at present only good enough for parametric statements. In  cases like the Wilson loops we get rid of the unknown constants through ratios. A curious situation prevails: we are indirectly witness of  the magnetic quasi-particles, but they remain for the time being the Poltergeists of the magnetic sector.   
Of course they are only approximately  free:  their density and pressure  are subject to a logarithmic rescaling of the temperature $g^2(T)T$, and hence at asymptotic temperatures the Stefan Boltzmann limit prevails.

\section{Suggestions for further reading}
Textbooks:\\
\begin{itemize}
\item {J. Kapusta, Finite Temperature Field Theory, CUP 1989.\\}
\item{M. Lebellac, Thermal field Theory, CUP 1996.\\}
\end{itemize}
Schools:\\
\begin{itemize}
\item M. Shaposhnikov (Erice 1996), High Temperature Effective Field Theory.\\

\item M. Laine (Trieste 2002), Finite Temperature Field Theory.\\
\item F. Karsch, Lect.Notes Phys. 583 (2002) 209-249, Lattice QCD at High Temperature and Density.\\
\item K.Kanaya, An introduction to finite temperature QCD on the lattice, 
 Prog.Theor.Phys.Suppl.131; 73, 1998; hep-lat/9804006.\\ 
\item Robert D. Pisarski, Notes on the Deconfining Phasetransition. 
 Proceedings of Cargese Summer School on QCD Perspectives on Hot and Dense Matter, Cargese, France, 6-18 Aug 2001; hep-ph/0203271.
\item A.K. Rebhan, Hard thermal loops and QCD thermodynamics.  Proceedings of Cargese Summer School on QCD Perspectives on Hot and Dense Matter, Cargese, France, 6-18 Aug 2001; hep-ph/0111341.
 \end{itemize}

%%%%%%%%%%%%%%%%%%%%%%%%%%%%%%%%%%figure%%%%%%%%%%%%%%%%%%%%%%%%%%%%%%%%%
%\begin{figure}
%\centerline{
%\includegraphics[viewport = 0 160 540 550,scale=0.44]{g.ps} %}
%\put(218,18){$\vec k^2/\omega_{\rm pl}^2$}
%\put(70,157){$\omega^2/\omega_{\rm pl}^2$}
%\bp\scriptsize 

%\ep
%\caption{The location of the zeros of $\Delta_t^{{\rm HTL}-1}$ (spatially
%transverse gauge bosons) 
%and of $\Delta_\ell^{{\rm HTL}-1}$ (longitudinal plasmons) in quadratic
%scales such as to show propagating modes and screening phenomena
%on one plot. }
%\label{Figg}
%\end{figure}
%%%%%%%%%%%%%%%%%%%%%%%%%%%%%%%%%%%%%%%%%%%%%%%%%%%%%%%%%%%%%%%%%%%%%%%%%

\section*{Acknowledgements}
My thanks are due to Prof. Zichichi and Prof. 't Hooft for their invitation
to this school.
For discussions on the subject I am indebted to Dietrich Boedeker, Philippe de Forcrand,
Pierre Giovannangeli, Christian Hoelbling, Oliver Jahn, Frithjof Karsch, Alex Kovner, Mikko Laine, Robert Pisarski, Tony Rebhan, Dan Son, Misha Stephanov and Michael Teper.

\section*{Appendix A: Free energy of heavy source and Wilson line average}

Below we prove eq.(\ref{eq:excess}) in the main text. It relates the free energy of the gauge averaged heavy quark state to the thermal average of the Wilson line.

 Start with the gauge average of a state with one heavy quark. The state with one quark, and eigenstate of the field operator $\vec A$
is 
\begin{equation}
\psi^{\dagger}_i|\vec A\rangle.
\label{eq:onefermi}
\end{equation}
We have suppressed reference to the position $\vec x$ of the quark.
Gauss' operator is $G(A_0)=Tr\vec E.\vec D A_0 +\psi^{\dagger}A_0\psi$ 
for an infinitesimal transformation. 
$A_0$ is dimensionless and is up to a dimensionful multiplicative  constant the scalar
potential at Euclidean time $\tau$, as in eq.(\ref{eq:wilsonlineordering}).
Integration over three-space is understood.
To render the state, eq.(\ref{eq:onefermi}), gauge invariant, we have to take the gauge projector $P_G\equiv \int DA_0\exp{iG(A_0)}$ and act with it on the state.
So the free energy excess due to the quark becomes:
\begin{equation}
\exp{-F_{\psi}}={1\over N}\sum^N_{i=1}\mbox{Tr}_{\vec A}\langle\vec A|\psi_iP_G\exp{-H/T}P_G\psi^{\dagger}_i|\vec A\rangle/\mbox{Tr}_{\vec A}\langle\vec A|P_G\exp{-H/T}P_G|\vec A\rangle
\label{excess}
\end{equation}

The gauge projector and the Hamiltonian commute. As in the usual transcription of the free energy into path integral language, we write for a partition of the Euclidean time interval into $n$ bits $\delta\tau={1\over{nT}}$:
\begin{equation}
P_G\exp{-H/T}P_G=P_G\exp{-\delta\tau H}P_G \exp{-\delta\tau HP_G}\ldots{...}P_G\exp{-\delta\tau H}P_G
\end{equation}
using  that $P_G$ is a projector. The transcription is identical to that of the free energy except that after rewriting the $n$ Hamiltonian factors 
as a path integral we are still left with the $n$ fermionic projectors $\exp{i\psi^{\dagger}A_0\psi}$. Every single one of them has the effect - through the canonical anticommutation relations-:
\begin{equation} 
\exp{(i\psi^{\dagger}A_0\psi)}\psi_i^{\dagger}|\vec A\rangle=\psi_k^{\dagger}(\exp{iA_0})_{ki}|\vec A\rangle.
\label{eq:fermiongaugeinv}
\end{equation}
 The reader will recognize the string bit appearing in eq.(\ref{eq:wilsonlineordering}) Repeat the operation in eq.(\ref{eq:fermiongaugeinv}) n times
and the  net result is the familiar time ordered exponential ${\cal{P}}(A_0)$ multiplied on the left by the fermion operators:
$$\psi_i\psi^{\dagger}_k({\cal{P}}(A_0))_{ki}.$$
Using once more the anti-commutation relations, one finds the anticipated relation between the thermal average of the Wilson line and the free energy excess:
\begin{equation}
\exp{-F_{\psi}}=\int DA\exp{-{1\over {g^2}}S(A)}{1\over N}\mbox{Tr}{\cal{P}}(A_0)/\int DA\exp{-{1\over {g^2}}S(A)}\equiv\langle P(A_0)\rangle.
\end{equation}
as in the main text, eq.(\ref{eq:excess}).

\noindent{\bf{Corollary}}\\
What is true for one heavy quark can  be shown along the same lines to be true for the correlator of heavy quarks: it is given by the correlator of two Wilson lines.


\begin{thebibliography}{xx}
\bibitem{ullie}For a review see: R. D. Pisarski, nucl-th/0212015; to appear in  Proceedings Quark Matter  2002, Nantes.

\bibitem{supercond}D.  Bailin, A. Love; Physics. Rept. 107,325, (1984); M. G. Alford,  K. Rajagopal, F. Wilczek, Nucl.Phys.B537:443-458,1999; hep-ph/9804403.

\bibitem{polyakov} A.M. Polyakov, Phys.Lett.B72,477 (1978); L. Susskind, Phys.Rev.D20,2610 (1979).

\bibitem{kanaya}K. Kanaya, An introduction to finite temperature QCD on the lattice,  Prog.Theor.Phys.Suppl.131:73-105,1998; hep-lat/9804006.

\bibitem{hands}S. Hands, Contemp.Phys.42:209, 2001. 

\bibitem{karschlect}F. Karsch, Lect.Notes Phys. 583 (2002) 209-249, Lattice QCD at High Temperature and Density.

\bibitem{lainechemical}P. de Forcrand, O. Philipsen, Nucl.Phys.B642, 290, 2002;   hep-lat/0205016. 

\bibitem{thooft}G.'t Hooft,  Nucl.Phys.B138, 1 (1978); 
Nucl.Phys.B153:141,1979 .

\bibitem{textbook}J. Kapusta, Finite Temperature Field Theory, Cambridge, 1989. M. Lebellac, Thermal Field Theory, CUP (1996).

\bibitem{shaposh}M. Shaposhnikov, Erice 1996, Effective theories and fundamental interactions, 360-383; hep-ph/9610247. 

\bibitem{tH76} G.~'t Hooft, in High Energy Physics, ed. A. Zichichi (Editrice
Compositori Bologna, 1976);
S.~Mandelstam, Phys.~Rep.~23C (1976), 245
T.~G.~Kov\'acs and E.~T.~Tomboulis, Phys.~Rev.~D57 (1998) 4054;
 T.~G.~Kov\'acs and E.~T.~Tomboulis, Phys.~Lett.~B463 (1999) 104;
J.~M.~Cornwall, Phys.~Rev.~D57 (1998) 7589;
J.~M.~Cornwall, Phys.~Rev.~D58 (1998) 1250;

 R.~Bertle, M.~Faber, J.~Greensite, S.~Olejnik,
  Nucl.~Phys.~Proc.Suppl. 83 (2000) 425;
 M.~Engelhardt, K.~Langfeld, H.~Reinhardt and O.~Tennert,
Phys.~Lett.~B431 (1998) 141.



\bibitem{kajdebye} K. Kajantie, M. Laine, K. Rummukainen, M. Shaposhnikov, Phys.Rev.Lett.79, 3130 (1997); hep-ph/9704416.

\bibitem{kajeff} K. Kajantie, M. Laine, K. Rummukainen, M. Shaposhnikov, Nucl. Phys. B503,(1997),357; hep-ph/9704416.

\bibitem{chapman}S. Chapman, Phys.Rev.D50:5308-5313,1994; hep-ph/9407313. 

\bibitem{huang}S.Z.Huang, M. Lissia,Nucl.Phys.B438,54,1995.

\bibitem{pis} T. Applequist, R.D. Pisarski, Phys.Rev.D23,2305,(1981); P. Ginsparg, Nucl.Phys.B170,388, (1980).

\bibitem{giovannakorthals} P. Giovannangeli, C.P. Korthals Altes, Nucl.Phys.B608, 203 (2001), hep-ph/0102022;  hep-ph/0212298.

\bibitem{bronoff}S. Bronoff, Th\`ese, Facult\'e de Luminy, Universit\'e de la M\'editerran\'ee, 1998.

\bibitem{linde} A. D. Linde, Phys.Lett. B96, 289(1980).

\bibitem{log}K. Kajantie, M. Laine, K. Rummukainen and Y. Schroeder, hep-ph/0211321, computes the 6th order log and reviews the lower order coefficients.

\bibitem{hartowe} A. Hart, O. Philipsen, Nucl.Phys.B572:243-265,2000; hep-lat/9908041. 

\bibitem{hartlaine}A. Hart, M. Laine, O. Philipsen, Nucl.Phys.B586:443-474,2000; hep-ph/0004060.
 
\bibitem{karschtension} F. Karsch, E. Laermann, M. Lutgemeier, Phys.Lett.B346:94-98,1995, hep-lat/9411020.



\bibitem{kuti}J. Kuti, J. Polonyi, K. Szlachanyi, Phys.Lett.B98, 199 (1981);L. D. McLerran, B. Svetitsky, Phys. Lett.B98, 195 (1981); N. Weiss, Phys.Rev.D24, 475 (1981).

\bibitem{gavin} S. Gavin, A. Gocksch, R. D. Pisarski, Phys.RevD49, 3079 (1994)
\bibitem{wilczek}R. D. Pisarski, F. Wilczek, Phys.Rev.D29, 338 (1984)
\bibitem{instanton}G.'t Hooft, Phys.Rev.Lett.37, 8,1976.
\bibitem{luciniteper}B. Lucini, M. Teper, Phys.Rev.D64,105019,2001;  hep-lat/0107007.
\bibitem{karschadj}F. Karsch, M. Luetgemaier, Nucl.Phys.B550: 449-464, 1998; hep-lat/9812023; F. Karsch, 
 in: Copenhagen 1998, Strong and Electroweak Matter 101-111; hep-lat/9903031 
\bibitem{rebhan}A.K. Rebhan, Phys.Rev.D48, 3967,1993; hep-ph/9308232. 
%\bibitem{direnzo}F. di Renzo, E. Onofri, G. Marchesini, P. Marenzoni, Nucl.Phy%s.B426, 675 (1994); hep-lat/9405019.
\bibitem{pispis}R. D. Pisarski, hep-ph/0203271, to appear in the proceedings of Cargese Summer School on QCD Perspectives on Hot and Dense Matter, Cargese, France. 
%\bibitem{guy}G.D.~Moore,
%``Pressure of hot QCD at large N(f),''
%JHEP {0210} (2002) 055
%[hep-ph/0209190].
%%CITATION = HEP-PH 0209190;%%

\bibitem{braaten}E.~Braaten and A.~Nieto,
%``Effective field theory approach to high temperature thermodynamics,''
Phys.\ Rev.\ D {51} (1995) 6990
[hep-ph/9501375].
%%CITATION = HEP-PH 9501375;%%
E. Braaten and A. Nieto,
%``Free Energy of QCD at High Temperature,''
Phys.\ Rev.\ D 53 (1996) 3421 [hep-ph/9510408].
%%CITATION = HEP-PH 9510408;%%
\bibitem{boyd}
G.~Boyd {\it et al.}, %J.~Engels, F.~Karsch, 
%E.~Laermann, C.~Legeland, M.~L\"utgemeier and B.~Petersson,
%``Thermodynamics of SU(3) Lattice Gauge Theory,''
Nucl.\ Phys.\  {B 469} (1996) 419 [hep-lat/9602007];
%%CITATION = HEP-LAT 9602007;%%
%A.~Papa,
%``SU(3) thermodynamics on small lattices,''
Nucl.\ Phys.\  {B 478} (1996) 335 [hep-lat/9605004];
%%CITATION = HEP-LAT 9605004;%%
%
B.~Beinlich, F.~Karsch, E.~Laermann and A.~Peikert,
%``String tension and thermodynamics
% with tree level and tadpole improved  actions,''
Eur.\ Phys.\ J.\  {C 6} (1999) 133 [hep-lat/9707023];
%%CITATION = HEP-LAT 9707023;%%
%
M.~Okamoto {\it et al.}  [CP-PACS Collaboration],
%``Equation of state for pure SU(3)
% gauge theory with renormalization  group improved action,''
Phys.\ Rev.\  {D 60} (1999) 094510 [hep-lat/9905005].
%%CITATION = HEP-LAT 9905005;%%

\bibitem{firstorder}P. Bacilieri et al., Phys.Rev.Lett.61(1988),1545; F.R. Brown et al,ibid. 61 (1988) 2058; A. Ukawa, Nucl. Phys. Proc.Suppl.B17 (1990) 118, ibid. 53 (1997) 106; E. Laermann, ibid. 63 (1988) 114; F. Karsch, ibid. 83 (2000) 14.
\bibitem{ohta}S. Ohta, M. Wingate, Nucl. Phys. Proc. Suppl. B73 (1999), 435; ibid. 83 (2000) 381; Phys.Rev.D 63 (2001)094502; R. V. Gavai, hep-lat/0110054. 
\bibitem{teperhot}B. Lucini, M. Teper, U. Wenger, Phys.Lett.B545, 197 (2002) 
, hep-lat/0206029. 
\bibitem{karschdebye}%O.~Kaczmarek, F.~Karsch, E.~Laermann and M.~Lutgemeier,
%Phys.Rev.D62 ~{034021}~(2000);  hep-lat/ 9908010.
S. Datta, S. Gupta, Phys.Lett.B471, 382 (2000);hep-lat/9906023; hep-lat/0208001
\bibitem{bhatta}T. Bhattacharya, A. Gocksch, C.P. Korthals Altes, R. D. Pisarski, Nucl.Phys.B 383, (1992),497; Phys.Rev.Lett.66,998 (1991); C.P. Korthals Altes, Nucl.Phys.B420 (1994), 637.
\bibitem{heller}J. Fingberg, U. Heller, F. Karsch, Nucl. Phys.B 392,(1993), 493;hep-lat/9208021.~ S. Gupta, Phys. Rev.D64 (2001) 034507.
\bibitem{zhai} P. Arnold, C. Zhai, Phys.Rev.D51, 1906, 1995; hep-ph/9410360; C. Zhai, B. Kastening, Phys.Rev.D52, 7232, 1995; hep-th/9507380.
\bibitem{arnoldyaffe}P. Arnold, L. Yaffe, Phys.Rev. D52 (1995) 7208; hep-ph/9508280. 
\bibitem{petrov}D. Diakonov, V. Petrov, Phys.Lett.B224; 131, 1989;  J.Exp.Theor.Phys.92, 905, 2001;  hep-th/0008035.
\bibitem{kovner}C. P. Korthals Altes, A. Kovner, Phys.Rev.D62, 096008, 2000; hep-ph/0004052.
\bibitem{stephanov}C. P. Korthals Altes, A. Kovner, M. Stephanov, Phys.Lett.B469, 205, 1999; hep-ph/9909516.
\bibitem{groeneveld}A.~Ukawa, P.~Windey and A.~Guth, Phys.~Rev.~D21 (1980) 1013; J. Groeneveld, J. Jurkiewicz, C.P. Korthals Altes, Phys.Scripta 23, 1022, 1981
\bibitem{teper3d} M. Teper, Phys.Rev.D59, 014512, 1999; hep-lat/9804008. 
\bibitem{defor}Ph. de Forcrand, M.d'Elia, M. Pepe, Phys.Rev.Lett.86, 1438 (2001); hep-lat/0007034.
\bibitem{rebbi}C. Hoelbling, C. Rebbi, V.A. Rubakov, Phys.Rev.D63:034506,2001; hep-lat/0003010; C. Hoebling, in preparation.
\bibitem{farakos}K. Farakos, 
P. de Forcrand,  C.P. Korthals Altes, M. Laine, M. Vettorazzo, hep-ph/0207343, to appear in Nucl.Phys.B.
\bibitem{extradim}
C.P.~Korthals Altes and M.~Laine,
Phys. Lett.B 511 (2001) 269; hep-ph/0104031; for the T=0 case: V. M. Belyaev, I.I. Kogan, Mod. Phys. Lett. A7 (1992),117.
\bibitem{alex}A. Kovner, Int.J.Mod.Phys.A17:2113, 2002; hep-th/0211248 
\bibitem{hoelbling} C. Hoelbling et al., in preparation; U. Wenger, private communication.
\bibitem{meyer} H. Meyer, private communication.
\end{thebibliography}
\end{document}